\documentstyle[prl,eqsecnum,amsmath,amssymb,amsthm,aps,epsf,epsfig,psfig]{revtex}

\theoremstyle{remark}

\newcommand{\ib}{\int\frac{d\omega}{2\pi i}}

\newcommand{\dk}{\hat{D}_{k}}
\newcommand{\ak}{\hat{A}_{k}}
\newcommand{\bk}{\hat{B}_{k}}
\newcommand{\pk}{\frac{\partial}{\partial k}}
\newcommand{\px}{\frac{\partial}{\partial x}}
\newcommand{\pp}{p_{+}}
\newcommand{\prm}{p_{-}}
\newcommand{\gp}[1]{g_{+}(#1)}
\newcommand{\gm}[1]{g_{-}(#1)}
\newcommand{\btk}{\beta(k)}
\newcommand{\sg}[1]{\sigma_{#1}}
\newcommand{\om}[1]{\omega_{#1}}

\begin{document}
\bibliographystyle{/home/m/mx200/paper/nonlinear/fwm/jcp}
\draft
\title{
\large \bf
Hyperpolarizabilities for the one-dimensional infinite single-electron 
periodic systems:\\
I. Analytical solutions under dipole-dipole correlations 
}
\author{Shidong Jiang}
\address{
Department of Mathematical Sciences, New Jersey Institute of Technology,
Newark, NJ 07102
}
\author{Minzhong Xu\thanks{Author to whom correspondence should be addressed. 
Email: mx200@nyu.edu}}
\address{
Department of Chemistry, New York University, New York, NY 10003
        }
\date{\today}
\maketitle
\bigskip
\begin{abstract}
The analytical solutions for the general-four-wave-mixing 
hyperpolarizabilities $\chi^{(3)}(-(\omega_1+\omega_2+\omega_3);\omega_1,
\omega_2,\omega_3)$
on infinite chains under both Su-Shrieffer-Heeger and 
Takayama-Lin-Liu-Maki models of trans-polyacetylene are obtained through
the scheme of dipole-dipole correlation. 
Analytical expressions of DC Kerr effect $\chi^{(3)}(-\omega;0,0,\omega)$, 
DC-induced second harmonic generation $\chi^{(3)}(-2\omega;0,\omega,\omega)$,
optical Kerr effect $\chi^{(3)}(-\omega;\omega,-\omega,\omega)$
and DC-electric-field-induced optical 
rectification $\chi^{(3)}(0;\omega,-\omega,0)$ are derived. 
By including or excluding ${\bf \nabla_k}$ terms in the calculations, 
comparisons show that the intraband contributions dominate the
hyperpolarizabilities if they are included. $\nabla_k$ term or intraband
transition leads to the break of the overall permutation symmetry 
in $\chi^{(3)}$
even for the low frequency and non-resonant regions. Hence it breaks the 
Kleinman symmetry that 
is directly based on the overall permutation symmetry. Our calculations 
provide a clear understanding of the Kleinman symmetry breaks that are 
widely observed in many experiments. We also suggest a feasible 
experiment on $\chi^{(3)}$ 
to test the validity of overall permutation symmetry
and our theoretical prediction. Finally, our calculations show the 
following trends for the various third-order nonlinear optical processes
in the low frequency and non-resonant region:
$\chi^{(3)}_{non-res}(-3\omega;\omega,\omega,\omega)>
\chi^{(3)}_{non-res}(-2\omega;0,\omega,\omega)>
\chi^{(3)}_{non-res}(-\omega;\omega,-\omega,\omega)
>\chi^{(3)}_{non-res}(-\omega;0,0,\omega) \ge
\chi^{(3)}_{non-res}(0;\omega,-\omega,0)$,
and in the resonant region:
$\chi^{(3)}_{res}(-\omega;0,0,\omega)>
\chi^{(3)}_{res}(-\omega;\omega,-\omega,\omega)>
\chi^{(3)}_{res}(-2\omega;0,\omega,\omega)
>\chi^{(3)}_{res}(0;\omega;-\omega,0)>
\chi^{(3)}_{res}(-3\omega;\omega,\omega,\omega)$. \\
\end{abstract}



\section{introduction}
\label{sec:intro}
The nonlinear optical (NLO) properties of $\pi$-conjugated polymers have been 
extensively studied by both experimentalists\cite{christ75,sauteret76,
chance80,greene87,kajzar89,halvorson92,krausz89,lawrence94,ledoux99,drenser99,kuebler00} and 
theorists\cite{agrawal,mxu1,mxu2,wwu,yuri,soos1,heflin,su1,su2,cwu1,cwu2,
shuai,mazumdar,soos2,zhang,karna,otto1,kirtman}.
Among these polymers, polyacetylene (PA) is the simplest conjugated polymer
and has been extensively studied\cite{heeger88}. PA consists of chains of 
CH units that form a pseudo one-dimensional(1D) lattice. 
Classical periodic single electron models like 
Su-Shrieffer-Heeger (SSH)\cite{ssh} and Takayama-Lin-Liu-Maki (TLM)\cite{tlm} 
have been established to interpret the optical properties of polyacetylene.
Recent experiments have measured the spectrum of third-order harmonic 
generations (THG) in polyacetylene\cite{kajzar89,halvorson92}, which requires
theoretical explanations. 
The contribution from the lattice relaxation to the third-order optical 
susceptibilities ($\chi^{(3)}$) is believed to be small because of the slow
process involved in the lattice relaxation\cite{kajzar87}.
Based on either the periodic single electron models 
(SSH and TLM) or the inclusion of electron-electron interactions extended
from both models, the frequency 
dependence of the THG in PA has been extensively studied 
theoretically\cite{agrawal,mxu1,mxu2,wwu,yuri,soos1,heflin,su1,su2,cwu1,cwu2,
shuai,mazumdar,soos2,zhang}.

Even for these simple single electron models, the theoretical studies on 
frequency dependence of $\chi^{(3)}$ have been approached in various 
ways: (i). Based on the different gauges, this problem can be studied either 
from dipole-dipole (DD)\cite{agrawal,mxu1,mxu2,su1,su2,cwu1,cwu2,shuai} or from 
current-current (JJ) correlations\cite{wwu,yuri}; (ii). Based on the size of 
chain, this problem can be treated either as a finite periodic chain
\cite{heflin,su1,su2,shuai,mazumdar,soos2} or an infinite one 
\cite{agrawal,mxu1,mxu2,wwu,cwu1}. Furthermore, from the point of view of different 
techniques to utilize the perturbative method, this problem can be 
solved either by quantum 
mechanics\cite{agrawal,soos1,heflin,su1,su2,cwu1,cwu2,shuai,mazumdar}
or by the field 
theory\cite{mxu1,mxu2,wwu,yuri}. Obviously, different approaches on the same
model should yield the same results if every step is carried out correctly.
However, this seemingly simple problem turns out to be nontrivial in the
actual computations because different results of $\chi^{(3)}$ for polyacetylene 
have been obtained from the different approaches. A typical example is the
THG calculation. There different gauges yield different results
both numerically and analytically\cite{mxu1,mxu2,wwu,su1}.
On the other hand, in the formulation of DD correlations, there are some 
controversies about whether or not the ${\bf \nabla_k}$ 
term should be included for an infinite periodic system\cite{otto1,kirtman,otto2,champagne,otto3}.
It is our interest to illustrate the differences and understand the reasons
why the discrepancies exist in the theoretical calculations.

Recently, many experiments showed the general failure of Kleinman 
symmetry\cite{kleinman} in describing the low frequency and 
non-resonant nonlinear optical properties of many materials and practical
systems\cite{dailey}. But most experiments only measured the $\chi^{(2)}$ 
in systems without centro- or inversion symmetry. Based on 
the possible mutual exclusion property between Krammers-Kronig dispersion 
relations and Kleinman symmetry, Dailey, Burke and Simpson asserted
the general failure of Kleinman symmetry. However, physically
Kleinman symmetry is only a direct consequence of the overall permutation 
symmetry\cite{kleinman,butcher}. 
Thus we would also like to investigate 
the overall permutation symmetry under SSH and TLM models in this work. 
For the trans-polyacetylene system where the centro- or inversion symmetry is
applied, $\chi^{(2)}$ vanishes. Hence we will use the
non-trivial 
results for $\chi^{(3)}$ to discuss the 
validity of both overall permutation and Kleinman symmetries.

In this and the subsequent\cite{jiang1} papers, we will present the analytical 
results of hyperpolarizabilities ($\chi^{(3)}$) for both the 
SSH and TLM models based on the 
field theory. Both SSH and TLM models describe periodic single electron
systems. The analytical form for the nonlinear optical response can be obtained
under both models and used to illustrate the subtle issues in periodic systems.
We will make detailed comparison between our calculation and other 
theoretical results of hyperpolarizabilities in both papers.
The purpose of present paper is to elucidate the physical contribution
of each term in the formulation of dipole-dipole correlations. In
these models, the important role played by the operator ${\bf \nabla_k}$
due to its noncommuting feature with ordinary functions of ${\bf k}$ is 
evident. Our calculations show that it is 
the ${\bf \nabla_k}$ term, which is physically related to the
intraband transition, that leads
to the break of the overall permutation symmetry and Kleinman symmetry
in the non-resonant region. Based on our calculations, we also suggest some 
experiments on the measurement of $\chi^{(3)}$ for certain physical systems
to test the validity of overall permutation symmetry
and our theoretical predictions.

This paper is organized as follows. In Section \ref{sec:theory},
necessary analytical tools and general theoretical framework are introduced.
In Section \ref{sec:hyper}, analytical 
results of four-wave-mixing (FWM), 
DC Kerr effect (DCKerr), DC-induced  second harmonic generation(DCSHG), 
optical Kerr effect or intensity-dependent index of refraction (IDIR), 
DC-electric-field-induced optical rectification (EFIOR) are derived.
The discussions
on those results and symmetries, comparisons with other theories
and some suggested experiments are then followed
in Section \ref{sec:discu}. A brief conclusion is presented in
Section \ref{sec:concl}. Finally, the details of analytical
calculations are 
given in 
the Appendix.

\section{Theory}
\label{sec:theory}
\subsection{Models and dipole operator}
Both SSH (or dimerized Huckel models) and TLM models have been thoroughly
studied in the {\it trans}-polyacetylene problems\cite{heeger88}. The SSH model
is an infinite 1D periodic model, and described by the following 
Hamiltonian\cite{ssh}:
\begin{eqnarray}
H_{SSH}=-\sum_{l,s} \left[ t_0+(-1)^l \frac{\Delta}{2} \right]
(\hat{C}_{l+1,s}^{\dag}\hat{C}_{l,s}^{}+\hat{C}_{l,s}^{\dag}\hat{C}_{l+1,s})^{},
\label{eq:Hssh}
\end{eqnarray}
where $t_0$ is the transfer integral between the nearest-neighbor sites,
$\Delta$ is the gap parameter and $\hat{C}_{l,s}^{\dag}(\hat{C}_{l,s})$
creates(annihilates) an $\pi$ electron at site $l$ with spin $s$. 
For the SSH model, each site is occupied by one electron.
In the continuum limit, the SSH model will tend to the TLM model given
by the formula \cite{tlm}:
\begin{eqnarray}
H_{TLM}=\Psi^{\dag}(x)\left(i\sigma_3v_F\partial_x+\Delta\sigma_1\right)
\Psi^{\dag}(x),
\label{eq:Htlm}
\end{eqnarray}
where $\Psi^{\dag}(x)=(\Psi^{\dag}_1(x),\Psi^{\dag}_2(x))$ is the 
two-component spinor describing the left-going and right-going electrons,
$v_F$ is the Fermi velocity, $\Delta$ is the gap parameter, and 
$\sigma_i$ ($i=1,2,3$) are the Pauli matrices.

Under the $DD$ correlation, the interaction Hamiltonian is expressed
by the formula
\begin{eqnarray}
\hat{H}_{\bf E\cdot r}=-e{\bf E\cdot \hat{r}}=-{\bf D\cdot E}, \nonumber
\end{eqnarray}
with $e$ the electron charge and ${\bf E}$ the electric field described as
follows:
\begin{eqnarray}
{\bf E}({\bf r}, t)={\bf E_0} e^{i{\bf k\cdot r}-i\om{} t},
\label{eq:e}
\end{eqnarray}
where ${\bf E_0}$ is the amplitude,  ${\bf k}$ and $\om{}$ are the
wave vector and frequency, respectively.

For periodic systems, the position operator ${\bf r}$ is often conveniently
defined in the momentum space\cite{blount}:
\begin{eqnarray}
{\bf r}_{n {\bf k}, n' {\bf k'}}= i \delta_{n,n'}{\bf \nabla_{k}}
\delta({\bf k}-{\bf k'}) + \Omega_{n,n'}({\bf k})\delta({\bf k}-{\bf k'}),
\label{eq:r}
\end{eqnarray}
where
\begin{eqnarray}
\displaystyle \Omega_{n,n'}({\bf k})=\frac{i}{v}\int_{v}
u_{n,{\bf k}}^*({\bf r}){\bf \nabla_{k}} u_{n', {\bf k}}({\bf r}) d {\bf r},
\end{eqnarray}
with $v$ the unit cell volume, $u_{n, {\bf k}}({\bf r})$  the periodic
function 
under the translation of lattice vector\cite{callaway}. Obviously, 
$u_{n, {\bf k}}({\bf r})$  is related to the wavefunction $\psi$ of
Bloch states by the formula:
\begin{eqnarray}
\psi_{n, {\bf k}}({\bf r}) = u_{n, {\bf k}}({\bf r})e^{i{\bf k \cdot r}},
\label{eq:bloch}
\end{eqnarray} 
where $n$ and ${\bf k}$ are the band index and crystal momentum, respectively. 

The energy bands of 1D SSH model are simple -- conduction and 
valence bands. Thus, this 1D model avoids certain problems due to the discontinuous
and non-analytical feature of Bloch wavefunctions 
for a composite band in higher-dimension periodic models\cite{zak1}.

Following the same procedures described in previous 
work\cite{mxu1,mxu2,mxu3}, we consider the momentum space
representation of the Hamiltonian given by Eq.\eqref{eq:Hssh}. With
the aid of the spinor description 
$\hat{\psi}_{k,s}^{\dag}(t)$=$(\hat{a}^{\dag c}_{k,s}(t)$, 
$\hat{a}^{\dag v}_{k,s}(t))$, where $\hat{a}^{\dag c}_{k,s}(t)$ and 
$\hat{a}^{\dag v}_{k,s}(t)$ are excitations of electrons in the conduction 
band and the valence band with momentum ${\bf k}$ and spin $s$, we obtain the 
following formula:
\begin{eqnarray}
\hat{H}_{SSH}(k,t)&=&\hat{H}_0+\hat{H}_{\bf E \cdot r} \\
&=&\sum_{-\frac{\pi}{2a}\le k\le\frac{\pi}{2a},s} \varepsilon(k) 
\hat{\psi}_{k,s}^{\dag}(t) \sg{3} \hat{\psi}_{k,s}(t) 
- \hat{D} \cdot E_0 e^{i\om{} t},
\label{eq:Hsshk}
\end{eqnarray}
where
$\vec{\sg{}}$ are the Pauli matrices, the operator $\hat{D}$ and the
parameter
$\varepsilon(k)$ are given by the following formulas, respectively.
\begin{eqnarray}
\hat{D}= e \sum_{-\frac{\pi}{2a}\le k\le\frac{\pi}{2a},s}
(\btk\, \hat{\psi}_{k,s}^{\dag} \sg{2}\hat{\psi}_{k,s}
 +i \frac{\partial}{\partial k} \, \hat{\psi}_{k,s}^{\dag}\hat{\psi}_{k,s}),
\label{eq:D}
\end{eqnarray}
\begin{eqnarray}
\varepsilon (k)= \sqrt{\left[ 2 t_0 cos(ka) \right]^2+\left[ \Delta sin(ka)
\right]^2},
\label{eq:ek}
\end{eqnarray}
The coefficient $\btk$ in Eq. \eqref{eq:D} is given by the formula:
\begin{eqnarray}
\btk=-\displaystyle \frac{\Delta t_0 a }{ \varepsilon^2(k)}.
\label{eq:be}
\end{eqnarray}
The coefficient $\btk$ is related to the interband transition 
between the conduction
and valence bands in a unit cell of length $2a$, and the second term in
Eq.\eqref{eq:D} is often related to the intraband transition\cite{mxu1,mxu2}.

\subsection{Berry phase and analytical format of one-dimensional Bloch 
functions}
In periodic systems, Bloch functions can not be analytical and are
discontinuous
for a composite band in which one band structure contains more than one 
branch. This fact was firstly pointed out by E.I. Blount\cite{blount}, 
and later
proved by J. Zak\cite{zak1}. Generally speaking, Eq.\eqref{eq:r} is not 
analytical
for composite bands. Fortunately, for the 1D periodic system being
discussed here,
it has been proved by Kohn that analytical results can be
obtained\cite{kohn} since
both valence and conducting bands are simple. Thus, we may avoid
the trouble due to the discontinuous or non-analytical feature of
Bloch wave functions.

The Berry phase for a 1D crystal with the centro- or inversion 
symmetric only 
can only be 0 or $\pi$ (mod $2\pi$)\cite{zak2}, therefore one can expects a
vanishing Berry phase for closed path in this specific problem. The Berry-phase
in the crystals is related to the diagonal matrix of first term in 
Eq.\eqref{eq:D}\cite{zak2,vanderbilt}. However, in the problem we are 
discussing, the Berry phase is 0.

\subsection{Field theory of hyperpolarizabilities}
There are a lot of different formulas to compute the hyperpolarizabilities 
-- Orr-Ward sum-over-state (SOS) method\cite{orr}, dipole formulas by 
Shen\cite{shen}, and Genkin-Mednis approach\cite{genkin} generalized by 
Bishop and his coauthors\cite{kirtman}. Different gauge approaches can be 
applied to this problem. All are based on the perturbation expansion. 
Recently, the field theory has been used to discuss this nonlinear 
problem\cite{mxu1,mxu2,wwu}.

For this single electron problem, the Feynman diagram of $\chi^{(3)}$ is simply 
described as one connected cycle in Fig.\ref{gr:x3circle}. If the 
long-wavelength approximation is applied, 
${\bf k_1}={\bf k_2}={\bf k_3}={\bf K}=0$. The only permutation to be
considered in this graph is three different frequencies $\om{1}$, $\om{2}$
and $\om{3}$. 

The general third-order susceptibility under ${\bf D\cdot E}$ gauge is described
by:
\begin{eqnarray}
&\chi&^{(3)}(\Omega; \om{1}, \om{2}, \om{3})= \frac{1}{3!V}
\left[ \frac{i}{\hbar} \right]^3 \int  d{\bf r}_1 d{\bf r}_2 d{\bf r}_3
\int dt_1 dt_2 dt_3 \nonumber \\
& &\int d{\bf r} dt\, e^{-i {\bf K \cdot r}+ i \Omega t} \langle \hat{T}
\hat{{\bf D}} ({\bf r},t) \hat{{\bf D}}({\bf r}_1,t_1)
\hat{{\bf D}}({\bf r}_2,t_2) \hat{{\bf D}} ({\bf r}_3,t_3) \rangle,
\label{eq:DD}
\end{eqnarray}
where $V$ is the total volume, $\displaystyle \Omega \equiv -\sum_{i=1}^{3}
\om{i}$, T is the time-ordering operator,  $\hat{{\bf D}}$ is the
dipole operator, and $\langle\cdots\rangle$ represents the average over the
unperturbed ground state.

For periodic systems, in order to maintain the periodicity of the
position operator ${\bf r}$, 
a "saw-liked" position operator must be introduced\cite{peeters}. 
For convenience in studying the nonlinear susceptibilities, we usually 
express Eqs.\eqref{eq:DD} in the momentum space
for further calculations\cite{mxu1,mxu2,mxu3}.

\section{Hyperpolarizabilities for SSH and TLM models under dipole formula}
\label{sec:hyper}
The TLM model\cite{tlm} is simply the continuum limit of the SSH
model\cite{ssh}, and  
analytical results for the TLM model can be easily derived from those under the SSH 
model. Therefore in
the following part, we will first focus on the hyperpolarizabilities under
 the SSH model.
Then we will deduce results under TLM model by simply passing to the
 continuum limit. As for the notations 
$\chi^{A}_{B}$, the superscript $A$ represents the abbreviation for the 
different four-wave-mixing (FWM) terms, the subscript $B$ represents the 
different models such as SSH or TLM models. We also use $\tilde{\chi}$ to 
represent the hyperpolarizabilities without considering $\nabla_k$ term 
in Eq.\eqref{eq:fwm1}.

\subsection{General four-wave-mixing(FWM) results}
\label{sec:hyper:fwm}
Under SSH and TLM models, the general four-wave-mixing(FWM) can be expressed as
$\chi_{SSH}^{FWM}(\om{1},\om{2},\om{3})$$\equiv$
$\chi_{SSH}^{(3)}(-(\om{1}+\om{2}+\om{3}),\om{1},\om{2},\om{3})$:
\begin{eqnarray}
\chi_{SSH}^{FWM}(\om{1},\om{2},\om{3})= \frac{2 e^4 n_0}{\hbar^3} 
\frac{1}{3!L}\sum_{k,{\mathcal{P}}(\om{1},\om{2},\om{3})}\int
\frac{id\om{}}{2\pi}Tr\Biggl\{ (&\beta&(k)\sg{2}+i\frac{\partial}{\partial k})
G(k,\om{}) \nonumber \\
(&\beta&(k)\sg{2}+i\frac{\partial}{\partial k}) G(k,\om{}-\om{1})
\nonumber \\
(&\beta&(k)\sg{2}+i\frac{\partial}{\partial k}) G(k,\om{}-\om{1}-\om{2})
\nonumber \\
(&\beta&(k)\sg{2}+i\frac{\partial}{\partial k}) G(k,\om{}-\om{1}-\om{2}-\om{3})
\Biggr\},
\label{eq:fwm1}
\end{eqnarray}
where $L$ is the chain length, $n_0$
is the number of chains per unit cross area, and 
${\mathcal{P}}(\om{1},\om{2},\om{3})$ represents all permutations for 
$\om{1}$, $\om{2}$ and $\om{3}$ (therefore the intrinsic symmetry is 
maintained\cite{butcher}). The polymer chains are assumed to
be oriented, and Green's function $G(k,\om{})$ is defined as
follows\cite{mxu1,mxu2,mxu3}:
\begin{eqnarray}
\displaystyle G(k,\om{})=
\frac{\om{}+\om{k}\sg{3}}{\om{}^2-\om{k}^2+i\epsilon},
\label{eq:green}
\end{eqnarray}
with $\om{k} \equiv \varepsilon(k) / \hbar \text{ and } \epsilon \equiv 0^+$.

After tedious derivations in the Appendix \ref{ap:fwm}, we obtain the 
following analytical results for the SSH model:
\begin{equation}\label{eq:FWMSSH}
\begin{aligned}
\chi^{FWM}_{SSH}(\om{1},\om{2},\om{3})
&=\chi_{0}^{(3)} \frac{15}{1024} \int_{1}^{1/\delta}\frac{xdx}
{\sqrt{(1-\delta^{2}x^2)(x^2-1)}} \\
&\left\{-\frac{(2x-z_{1}-z_{3})}
{x^{8}(x-z_{1})(x+z_{2})(x-z_{3})(x-z_{1}-z_{2}-z_{3})}\right.\\
&-\frac{(2x+z_{1}+z_{3})}
{x^{8}(x+z_{1})(x-z_{2})(x+z_{3})(x+z_{1}+z_{2}+z_{3})}\\
&+\frac{4(1-\delta^{2}x^2)(x^2-1)}{x^8(x-z_{1}-z_{2})}
\frac{(3x-2z_{1})(3x-2(z_{1}+z_{2}+z_{3}))}
{(x-z_{1})^{2}(x-z_{1}-z_{2}-z_{3})^2}\\
&+\left.\frac{4(1-\delta^{2}x^2)(x^2-1)}{x^8(x+z_{1}+z_{2})}
\frac{(3x+2z_{1})(3x+2(z_{1}+z_{2}+z_{3}))}
{(x+z_{1})^{2}(x+z_{1}+z_{2}+z_{3})^2}\right\},
\end{aligned}\end{equation}
where
\begin{equation}
\chi_{0}^{(3)}=\frac{8}{45}\frac{e^4n_{0}(2t_0a)^3}{\pi\Delta^6}
\end{equation}
and
\begin{equation}  \label{eq:zi}
z_{i}=\frac{\hbar\om{i}}{2\Delta}, \quad \text{for $i=1,2,3$}.
\end{equation}
By setting $z_1=z_2=z_3=z=\hbar \omega/2\Delta$, Eq.\eqref{eq:FWMSSH} 
can be simplified as third 
harmonic generation $\chi^{(3)}(-3\omega; \omega,\omega,\omega)$. It is easy to
prove that Eq.\eqref{eq:FWMSSH} is the same as Eq.(2.20)\cite{mxu1}
or Eq.(9)\cite{mxu2} in our previous works.

\begin{figure}
\centerline{
\epsfxsize=9cm \epsfbox{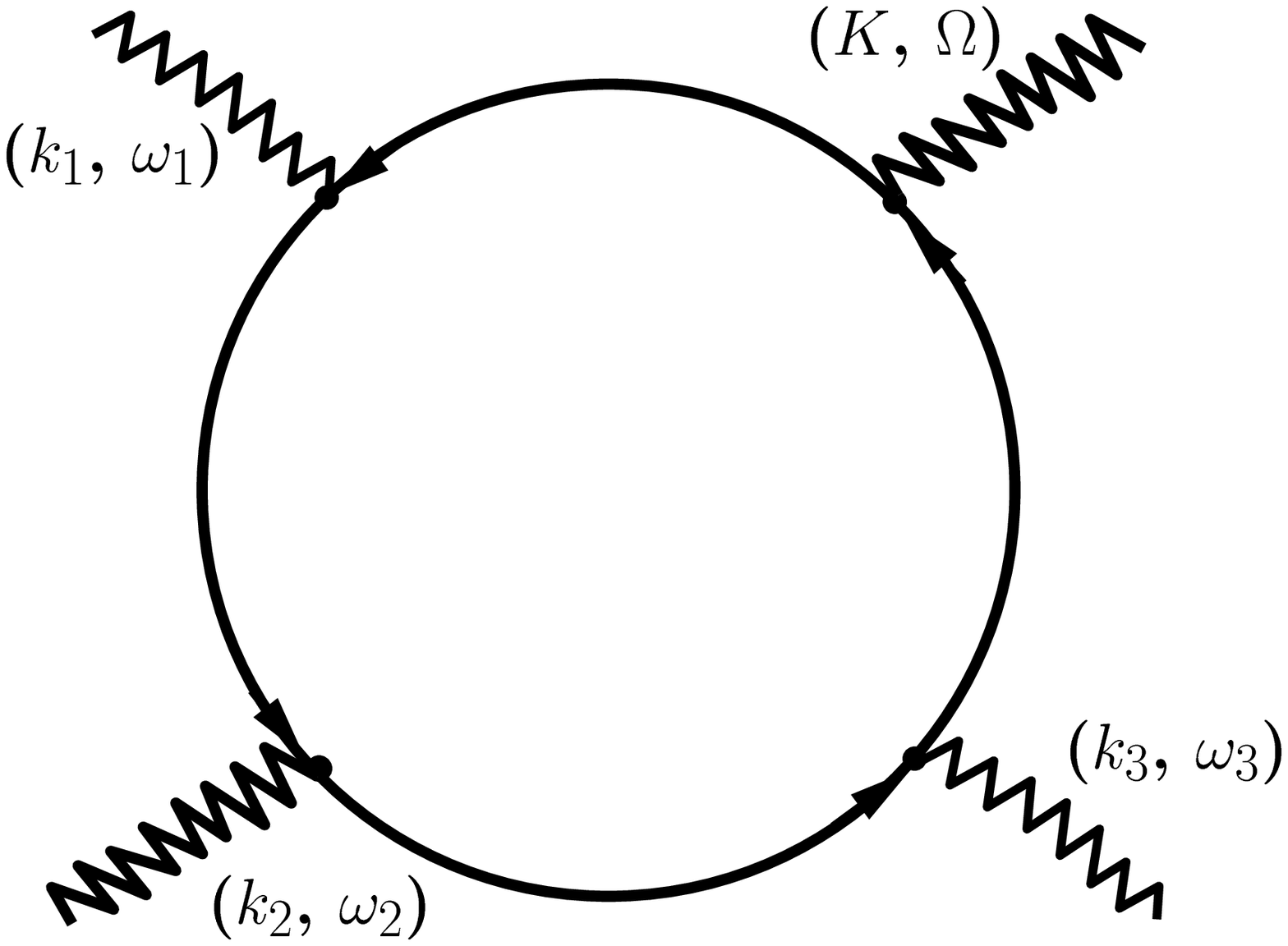}
}
\caption{The Feynman diagram for 
$\chi^{(3)}(-\Omega,{\bf -K};\om{1},{\bf k_1},\om{2},{\bf k_2},
\om{3},{\bf k_3})$, 
where $\Omega=\om{1}+\om{2}+\om{3}$, 
${\bf K}={\bf k_1}+{\bf k_2}+{\bf k_3}$.}
\label{gr:x3circle}
\end{figure}

By changing $x \to x+i\epsilon$ in Eq.\eqref{eq:FWMSSH}, and by choosing the 
same parameters used in our previous works 
for polyacetylene\cite{mxu1,mxu2,cwu1,cwu2}, $\Delta=0.9 eV$, 
$n_0=3.2 \times 10^{14} cm^{-2}$, $a=1.22 \AA$  and $\epsilon \sim 0.03$,
we have $\delta=0.18$ and $\chi_0^{(3)} \approx 1.0 \times 10^{-10}$ esu.
The absolute value of FWM is plotted in Fig.\ref{gr:x3circle}. 

From the graph, we find several symmetrical resonant frequencies. The biggest
resonant peaks are around $(0, 0, \pm 1)$, $(\pm 1, \pm 1, \mp 1)$ and their
permutations, which correspond to DC Kerr effect(DCKerr) and Optical Kerr 
effect or Intensity-dependent index of refraction(IDIR), respectively.
There are some secondary resonant frequencies, the cusps shown in the 
Fig.\ref{gr:x3circle}
are around $(\pm 1, \mp 1, 0)$ and their permutations. They correspond to 
DC-electric-field-induced optical rectification(EFIOR). The resonant peaks for 
third-harmonic generation (THG) are not obvious in this graph.

As to the magnitudes of resonant peaks, we have 
$\chi^{(3)}(-\omega;0,0,\omega)\gg\chi^{(3)}(-\omega;\omega,\omega,-\omega)
\gg\chi^{(3)}(0;\omega,-\omega,0)\gg \chi^{(3)}(-3\omega;\omega,\omega,\omega)$.

Eq.\eqref{eq:FWMSSH} could be further simplified 
as hyperpolarizabilities under TLM model by letting $\delta \rightarrow 0$: 
\begin{equation}\begin{split}
\chi^{FWM}_{TLM}(\om{1},\om{2},\om{3}) &= \frac{45\chi_{0}^{(3)}}{64}
\left\{\frac{Z^{6}}
{z_{1}z_{2}z_{3}\sigma}L(4,Z)-\frac{Z^{6}}{z_{1}^{2}z_{2}^{2}z_{3}^{2}}M(4,Z)
+\sum_{i=1}^{3}\frac{z_{i}^{3}}{\sigma}L(4,z_{i})\right.\\
&+\left.\sum_{P(z_1,z_2,z_3)}\frac{z_{1}^{2}(-z_{1}+2(z_{2}+z_{3}))}
{2z_{2}z_{3}(z_{2}+z_{3})}L(4,z_{1})\right\} \\
&+\frac{45\chi_{0}^{(3)}}{64}
\left\{\sum_{P(z_1,z_2,z_3)}\frac{(z_{1}+z_{2})^{5}(z_{1}+z_{2}-2z_{3})}
{2z_{1}^{2}z_{2}^{2}z_{3}^{2}}M(4,z_{1}+z_{2})\right.\\
&+\left.\frac{-z_{1}^{2}(z_{1}^{2}
-2z_{1}(z_{2}+z_{3})+6z_{2}z_{3})}
{2z_{2}^{2}z_{3}^{2}}M(4,z_{1})\right\},
\end{split}\end{equation}
where
\begin{equation}
\sigma := (z_{1}+z_{2})(z_{2}+z_{3})(z_{3}+z_{1}),
\end{equation}
\begin{equation}
Z := z_1+z_2+z_3,
\end{equation}
$L(n,z)$ and $M(n,z)$ $(n=0\ldots 4)$ are defined by
\eqref{eq:lnz} and \eqref{eq:mnz} respectively.

\subsection{DC Kerr effect}
By setting $z_1=z_2=0$ and $z_3=z$ ($z=\hbar\omega/(2\Delta)$) in Eq.\eqref{eq:chi},
 we have 

\begin{equation}
\begin{aligned}
\chi^{(3)}(0,0,\omega)
&=\chi_{0}^{(3)}\frac{15}{256}\int_{1}^{1/\delta}\frac{dx}
{x^7\sqrt{(1-\delta^{2}x^2)(x^2-1)}} \\
&\left\{-\frac{(2x-z)}{x^2(x-z)^2}-\frac{(2x+z)}{x^2(x+z)^2}
-\frac{2}{x(x^2-z^2)}\right.\\
&+2(1-\delta^{2}x^2)(x^2-1)\left\{\frac{3(3x-2z)}{x^2(x-z)^2}
+\frac{3(3x+2z)}{x^2(x+z)^2}\right.\\
&+\left.\left.\frac{(3x-2z)^2}{(x-z)^5}+\frac{(3x+2z)^2}{(x+z)^5}
+\frac{3(3x-2z)}{x(x-z)^3}+\frac{3(3x+2z)}{x(x+z)^3}\right\}
\right\}
\end{aligned}\end{equation}

After simplifications, we obtain the DC Kerr effect(DCKerr) coefficient under 
the SSH model: 
\begin{equation}   
\label{eq:dckerrssh}
\begin{aligned}
\chi_{SSH}^{DCKerr}(0,0,\omega)
&=\chi_{0}^{(3)}\frac{15}{128}\int_{1}^{1/\delta}\frac{dx}
{\sqrt{(1-\delta^{2}x^2)(x^2-1)}} \\
&\left\{-\frac{2}{x^6(x^2-z^2)^2}-\frac{1}{x^8(x^2-z^2)}\right.\\
&+2(1-\delta^{2}x^2)(x^2-1)
\left\{\frac{16}{(x^2-z^2)^5}+\frac{12}{x^2(x^2-z^2)^4}\right.\\
&+\left.\left.\frac{1}{x^4(x^2-z^2)^3}+\frac{1}{x^6(x^2-z^2)^2}-\frac{3}{x^8(x^2
-z^2)}\right\}\right\}
\end{aligned}\end{equation}

As $\delta \rightarrow 0$ in Eq.\eqref{eq:dckerrssh}, we have the results 
under TLM model:
\begin{equation}
\begin{aligned}
\chi^{DCKerr}_{TLM}(0,0,\omega)
&=\chi_{0}^{(3)}\frac{15}{128}\left(-2L(3,z)-L(4,z)+32M(0,z)\right.\\
&+\left.24M(1,z)+2M(2,z)+2M(3,z)-6M(4,z)\right),
\end{aligned}\end{equation}

where $L(n,z)$ and $M(n,z)$ $(n=0\ldots 4)$ are defined by
\eqref{eq:lnz} and \eqref{eq:mnz} respectively.

Substituting the identities for $L(n,z)$ and $M(n,z)$ into the above
equation and simplifying the resulting expression, we obtain
\begin{equation}
\label{eq:dckerrtlm}
\begin{aligned}
\chi^{DCKerr}_{TLM}(0,0,\omega)
&=\chi_{0}^{(3)}\frac{15}{512z^8(z^2-1)^3}\left\{
-(120z^8-580z^6+1029z^4-780z^2+216)f(z)\right.\\
&+\left.\frac{1}{105}(384z^{12}-928z^{10}+760z^8-22182z^6+65541z^4-66780z^2+2268
0)\right\}
\end{aligned}\end{equation}

Dropping the terms related to $\nabla_k$ or $\pk$ in Eq.\eqref{eq:fwm1}, we have
\begin{equation}
\label{eq:dkdckerrssh}
\begin{aligned}
\tilde{\chi}^{DCKerr}_{SSH}(0,0,\omega)
&=\chi_{0}^{(3)}\frac{15}{128}\int_{1}^{1/\delta}\frac{dx}
{\sqrt{(1-\delta^{2}x^2)(x^2-1)}}
\left\{-\frac{2}{x^6(x^2-z^2)^2}-\frac{1}{x^8(x^2-z^2)}\right\}
\end{aligned}\end{equation}

As $\delta \rightarrow 0$ in the Eq.\eqref{eq:dkdckerrssh}, we have
\begin{equation}
\begin{aligned}
\tilde{\chi}^{DCKerr}_{TLM}(0,0,\omega)
&=-\chi_{0}^{(3)}\frac{15}{128}\left(2L(3,z)+L(4,z)\right),
\end{aligned}\end{equation}

where $L(n,z)$ is defined by \eqref{eq:lnz}.

Substituting the identities for $L(n,z)$ into the above
equation and simplifying the resulting expression, we obtain
\begin{equation}
\label{eq:dkdckerrtlm}
\begin{aligned}
\tilde{\chi}^{DCKerr}_{TLM}(0,0,\omega)
&=\chi_{0}^{(3)}\frac{15}{128z^8(z^2-1)}\left\{
(7z^2-6)f(z)\right.\\
&+\left.\frac{1}{105}(48z^8-104z^6-154z^4-315z^2+630)\right\}
\end{aligned}\end{equation}

\subsection{DC induced second harmonic generation}
By setting $z_1=z_2=z$ ($z=\hbar\omega/(2\Delta)$) and $z_3=0$ in 
Eq.\eqref{eq:chi}, we have 
\begin{equation}
\begin{aligned}
\chi^{(3)}(0,\omega,\omega)
&=\frac{15}{256}\chi_{0}^{(3)}\int_{1}^{1/\delta}\frac{dx}
{x^7\sqrt{(1-\delta^{2}x^2)(x^2-1)}} \\
&\left\{-\frac{(2x-z)}{x(x^2-z^2)(x-2z)}-\frac{(2x+z)}{x(x^2-z^2)(x+2z)}\right
.\\
&-\frac{1}{x(x-z)(x-2z)}-\frac{1}{x(x+z)(x+2z)}\\
&+2(1-\delta^{2}x^2)(x^2-1)
\left\{\frac{3x(3x-4z)}{x^2(x-z)(x-2z)^2}
 +\frac{3x(3x+4z)}{x^2(x+z)(x+2z)^2}\right.\\
&+\frac{(3x-2z)(3x-4z)}{(x-z)^3(x-2z)^2}
 +\frac{(3x+2z)(3x+4z)}{(x+z)^3(x+2z)^2}\\
&+\left.\left.\frac{(3x-2z)(3x-4z)}{(x-z)^2(x-2z)^3}
 +\frac{(3x+2z)(3x+4z)}{(x+z)^2(x+2z)^3}\right\}
\right\}
\end{aligned}\end{equation}

After simplifications, we obtain the DC induced second harmonic generation(DCSHG)
coefficient under SSH model:

\begin{equation} 
\label{eq:dcshgssh}
\begin{aligned}
\chi^{DCSHG}_{SSH}(0,\omega,\omega)
&=\frac{15}{128}\chi_{0}^{(3)}\int_{1}^{1/\delta}\frac{dx}
{\sqrt{(1-\delta^{2}x^2)(x^2-1)}} \\
&\left\{-\frac{4}{x^8(x^2-4z^2)}+\frac{1}{x^8(x^2-z^2)}\right.\\
&+2(1-\delta^{2}x^2)(x^2-1)
\left\{\frac{6}{x^8(x^2-z^2)}+\frac{1}{x^6(x^2-z^2)^2}-\frac{4}{x^4(x^2-z^2)^3}
\right.\\
&-\left.\left.\frac{36}{x^8(x^2-4z^2)}+\frac{28}{x^6(x^2-4z^2)^2}
 +\frac{32}{x^4(x^2-4z^2)^3}\right\}
\right\}
\end{aligned}\end{equation}

As $\delta \rightarrow 0$ in Eq.\eqref{eq:dcshgssh}, we have the result under
the TLM model:
\begin{equation}
\begin{aligned}
\chi^{DCSHG}_{TLM}(0,\omega,\omega)
&=\frac{15}{128}\chi_{0}^{(3)}\left(-4L(4,2z)+L(4,z)+12M(4,z)+2M(3,z)\right.\\
&\left.-8M(2,z)-72M(4,2z)+56M(3,2z)+64M(2,2z)\right),
\end{aligned}\end{equation}

where $L(n,z)$ and $M(n,z)$ $(n=0\ldots 4)$ are defined by
\eqref{eq:lnz} and \eqref{eq:mnz} respectively.

Substituting the identities for $L(n,z)$ and $M(n,z)$ into the above
equation and simplifying the resulting expression, we obtain
\begin{equation}
\label{eq:dcshgtlm}
\begin{aligned}
\chi^{DCSHG}_{TLM}(0,\omega,\omega)
&=\chi_{0}^{(3)}\frac{15}{256z^8(z^2-1)^2}\left\{(28z^6-104z^4+118z^2-43)
f(z)\right.\\
&-\left.\frac{1}{105}(288z^{10}-184z^8+204z^6-5068z^4+9380z^2-4515)\right\}\\
&+\chi_{0}^{(3)}\frac{15}{2048z^8(4z^2-1)}\left\{(-48z^4+6z^2+1)f(2z)\right.\\
&+\left.\frac{1}{105}(12288z^8+4992z^6+2464z^4-910z^2-105)\right\}
\end{aligned}\end{equation}

Dropping the terms related to $\nabla_k$ or $\pk$ in Eq.\eqref{eq:fwm1}, we have
\begin{equation} 
\label{eq:dkdcshgssh}
\begin{aligned}
\tilde{\chi}^{DCSHG}_{SSH}(0,\omega,\omega)
&=\chi_{0}^{(3)}\frac{15}{128}\int_{1}^{1/\delta}\frac{dx}
{\sqrt{(1-\delta^{2}x^2)(x^2-1)}}
\left\{-\frac{4}{x^8(x^2-4z^2)}+\frac{1}{x^8(x^2-z^2)}
\right\}
\end{aligned}\end{equation}

As $\delta \rightarrow 0$ in Eq.\eqref{eq:dkdcshgssh}, we have
\begin{equation} 
\begin{aligned}
\tilde{\chi}^{DCSHG}_{TLM}(0,\omega,\omega)
&=\chi_{0}^{(3)}\frac{15}{128}\left(-4L(4,2z)+L(4,z)\right),
\end{aligned}\end{equation}

where $L(n,z)$ is defined by \eqref{eq:lnz}.

Substituting the identities for $L(n,z)$ into the above
equation and simplifying the resulting expression, we obtain
\begin{equation} 
\label{eq:dkdcshgtlm}
\begin{aligned}
\tilde{\chi}^{DCSHG}_{TLM}(0,\omega,\omega)
&=\chi_{0}^{(3)}\frac{15}{1024z^8}\left\{
8f(z)-\frac{1}{8}f(2z)-\frac{1}{40}(128z^4+200z^2+315)\right\}
\end{aligned}\end{equation}

\subsection{Optical Kerr effect or intensity-dependent index of refraction}
By setting $z_1=z_3=z$ ($z=\hbar\omega/(2\Delta)$) and $z_2=-z$ in 
Eq.\eqref{eq:chi}, we have
\begin{equation}
\begin{aligned}
\chi^{(3)}(\omega,-\omega,\omega)
&=\chi_{0}^{(3)}\frac{15}{256}\int_{1}^{1/\delta}\frac{dx}
{x^7\sqrt{(1-\delta^{2}x^2)(x^2-1)}} \\
&\left\{-\frac{4x}{(x^2-z^2)^2}-\frac{1}{(x-z)^3}-\frac{1}{(x+z)^3}\right.\\
&+2(1-\delta^{2}x^2)(x^2-1)
\left\{\frac{(3x-2z)^2}{x(x-z)^4}
 +\frac{(3x+2z)^2}{x(x+z)^4}\right.\\
&+\left.\left.\frac{(3x-2z)^2}{(x-2z)(x-z)^4}
 +\frac{(3x+2z)^2}{(x+2z)(x+z)^4}+\frac{2(9x^2-4z^2)}{x(x^2-z^2)^2}\right\}
\right\}
\end{aligned}\end{equation}

After simplifications,  we obtain the optical Kerr effect or intensity-dependent
index of refraction (IDIR) coefficient under the SSH model:
\begin{equation} 
\label{eq:idirssh}
\begin{aligned}
\chi^{IDIR}_{SSH}(\omega,-\omega,\omega)
&=\chi_{0}^{(3)}\frac{15}{128}\int_{1}^{1/\delta}\frac{dx}
{\sqrt{(1-\delta^{2}x^2)(x^2-1)}} \\
&\left\{\frac{1}{x^6(x^2-z^2)^2}-\frac{4}{x^4(x^2-z^2)^3}\right.\\
&+2(1-\delta^{2}x^2)(x^2-1)
\left\{-\frac{16}{x^8(x^2-z^2)}-\frac{13}{x^6(x^2-z^2)^2}\right.\\
&-\left.\left.\frac{8}{x^4(x^2-z^2)^3}+\frac{64}{x^8(x^2-4z^2)}\right\}
\right\}
\end{aligned}\end{equation}

As $\delta \rightarrow 0$ in Eq. \eqref{eq:idirssh}, we have the result 
under the TLM model:
\begin{equation} 
\begin{aligned}
\chi^{IDIR}_{TLM}(\omega,-\omega,\omega)
&=\chi_{0}^{(3)}\frac{15}{128}\left(L(3,z)-4L(2,z)-32M(4,z)\right.\\
&\left.-26M(3,z)-16M(2,z)+128M(4,2z)\right),
\end{aligned}\end{equation}

where $L(n,z)$ and $M(n,z)$ $(n=0\ldots 4)$ are defined by
\eqref{eq:lnz} and \eqref{eq:mnz} respectively.

Substituting the identities for $L(n,z)$ and $M(n,z)$ into the above
equation and simplifying the resulting expression, we obtain
\begin{equation} 
\label{eq:idirtlm}
\begin{aligned}
\chi^{IDIR}_{TLM}(\omega,-\omega,\omega)
&=\chi_{0}^{(3)}\frac{15}{256z^8}\left\{
\frac{8z^8+44z^6+83z^4-495z^2+315}{15(z^2-1)^2}\right.\\
&\left.-\frac{4z^6+22z^4-43z^2+20}{(z^2-1)^2}f(z)+(4z^2-1)f(2z)\right\}
\end{aligned}\end{equation}

Dropping the terms related to $\nabla_k$ or $\pk$ in Eq.\eqref{eq:fwm1}, we have
\begin{equation} 
\label{eq:dkidirssh}
\begin{aligned}
\tilde{\chi}^{IDIR}_{SSH}(\omega,-\omega,\omega)
&=\chi_{0}^{(3)}\frac{15}{128}\int_{1}^{1/\delta}\frac{dx}
{\sqrt{(1-\delta^{2}x^2)(x^2-1)}}
\left\{\frac{1}{x^6(x^2-z^2)^2}-\frac{4}{x^4(x^2-z^2)^3}\right\}
\end{aligned}\end{equation}

As $\delta \rightarrow 0$ in the Eq.\eqref{eq:dkidirssh}, we have
\begin{equation}
\begin{aligned}
\tilde{\chi}^{IDIR}_{TLM}(\omega,-\omega,\omega)
&=\chi_{0}^{(3)}\frac{15}{128}\left(L(3,z)-4L(2,z)\right),
\end{aligned}\end{equation}
where $L(n,z)$ is defined by \eqref{eq:lnz}.

Substituting the identities for $L(n,z)$ into the above
equation and simplifying the resulting expression, we obtain
\begin{equation} 
\label{eq:dkidirtlm}
\begin{aligned}
\tilde{\chi}^{IDIR}_{TLM}(0,\omega,\omega)
&=\chi_{0}^{(3)}\frac{1}{256z^8(z^2-1)^2}\left\{
-(840z^4-1425z^2+630)f(z)\right.\\
&\left.+(16z^8+88z^6+226z^4-1005z^2+630)\right\}
\end{aligned}\end{equation}

\subsection{DC-electric-field-induced optical rectification}
By setting $z_1=0$, $z_2=-z$, $z_3=z$ ($z=\hbar\omega/(2\Delta)$) 
in Eq.\eqref{eq:chi}, we obtain the 
DC-electric-field-induced optical rectification (EFIOR) under the SSH model:

\begin{equation} 
\label{eq:efiorssh}
\begin{aligned}
\chi^{EFIOR}_{SSH}(0,-\omega,\omega)
&=\chi_{0}^{(3)}\frac{15}{128}\int_{1}^{1/\delta}\frac{dx}
{x^8\sqrt{(1-\delta^{2}x^2)(x^2-1)}} \\
&\left\{\frac{-3x^2+z^2}{(x^2-z^2)^2}
+6(1-\delta^{2}x^2)(x^2-1)
\frac{2z^4+9x^4-7x^2z^2}{(x^2-z^2)^3}\right\}\\
&=\chi_{0}^{(3)}\frac{15}{128}\int_{1}^{1/\delta}\frac{dx}
{\sqrt{(1-\delta^{2}x^2)(x^2-1)}} \\
&\left\{\frac{-2}{x^6(x^2-z^2)^2}-\frac{1}{x^8(x^2-z^2)}\right.\\
&+\left.6(1-\delta^{2}x^2)(x^2-1)
\left\{\frac{4}{x^4(x^2-z^2)^3}+\frac{3}{x^6(x^2-z^2)^2}
+\frac{2}{x^8(x^2-z^2)}\right\}\right\}
\end{aligned}\end{equation}

As $\delta \rightarrow 0$, we have the result under the TLM model:
\begin{equation}
\begin{aligned}
\chi^{EFIOR}_{TLM}(0,-\omega,\omega)
&=\chi_{0}^{(3)}\frac{15}{128}
\left(-2L(3,z)-L(4,z)+6(4M(2,z)+3M(3,z)+2M(4,z))\right),
\end{aligned}\end{equation}

where $L(n,z)$ and $M(n,z)$ $(n=0\ldots 4)$ are defined by
\eqref{eq:lnz} and \eqref{eq:mnz} respectively.

Substituting the identities for $L(n,z)$ and $M(n,z)$ into the above
equation and simplifying the resulting expression, we obtain
\begin{equation} 
\label{eq:efiortlm}
\begin{aligned}
\chi^{EFIOR}_{TLM}(0,-\omega,\omega)
&=\chi_{0}^{(3)}\frac{15}{64z^8(z^2-1)}\left\{
-\frac{1}{105}(24z^8-38z^6+119z^4-2520z^2+2520)\right.\\
&+\left.(15z^4-40z^2+24)f(z)\right\}
\end{aligned}\end{equation}

Dropping the terms related to $\nabla_k$ or $\pk$ in Eq.\eqref{eq:fwm1}, 
we have
\begin{equation}
\begin{aligned}
\tilde{\chi}^{EFIOR}_{SSH}(0,-\omega,\omega)
=\chi_{0}^{(3)}\frac{15}{128}\int_{1}^{1/\delta}\frac{dx}
{\sqrt{(1-\delta^{2}x^2)(x^2-1)}}
\left\{\frac{-2}{x^6(x^2-z^2)^2}-\frac{1}{x^8(x^2-z^2)}\right\}
\label{eq:dkefiorssh}
\end{aligned}\end{equation}

As $\delta \rightarrow 0$ in the Eq.\eqref{eq:dkefiorssh}, we have
\begin{equation}
\begin{aligned}
\tilde{\chi}^{EFIOR}_{TLM}(0,-\omega,\omega)
=\chi_{0}^{(3)}\frac{15}{128}\left(-2L(3,z)-L(4,z)\right),
\end{aligned}\end{equation}
where $L(n,z)$ is defined by \eqref{eq:lnz}.

Substituting the identities for $L(n,z)$ and $M(n,z)$ into the above
equation and simplifying the resulting expression, we obtain
\begin{equation} 
\label{eq:dkefiortlm}
\begin{aligned}
\tilde{\chi}^{EFIOR}_{TLM}(0,-\omega,\omega)
&=\chi_{0}^{(3)}\frac{15}{128z^8(z^2-1)}\left\{
(7z^2-6)f(z)\right.\\
&+\left.\frac{1}{105}(48z^8-104z^6-154z^4-315z^2+630)\right\}
\end{aligned}\end{equation}

\subsection{Third harmonic generation}
The results of third harmonic generation(THG) can be computed by setting 
$z_1=z_2=z_3=z$ in Eq.\eqref{eq:chi3}. They can also be found in our previous 
works\cite{mxu1,mxu2}. Here we simply state the result:
\begin{eqnarray}
\chi_{SSH}^{THG}(\omega)
=\chi_0^{(3)} \frac{45}{128} \int_1^{\frac{1}{\delta}}
\frac{d x}{ \sqrt{(1-\delta^2 x^2)(x^2-1)}}
\Biggl\{
&-&\frac{47-48(1+\delta^2)x^2+48\delta^2x^4}{8 x^8(x^2-z^2)} \nonumber \\
&+&\frac{3(1-\delta^2 x^2)(x^2-1)}{x^6 (x^2-z^2)^2} \nonumber \\
&+&\frac{9\left[ 47-48(1+\delta^2)x^2+48\delta^2 x^4 \right]}
{8 x^8(x^2-(3 z)^2)}\nonumber \\
&+&\frac{63(1-\delta^2 x^2)(x^2-1)} {x^6 (x^2-(3 z)^2)^2} \Biggr\}
\label{eq:thgssh}
\end{eqnarray}
and
\begin{eqnarray}
\chi_{TLM}^{THG}(\omega)= \chi_0^{(3)} \frac{45}{128} \, \Biggl\{
-\frac{14}{3 z^8}-\frac{4}{15 z^4}
+\frac{(37-24 z^2)}{8 z^8} f(z)+\frac{(1-8 z^2)}{24 z^8} f(3z) \Biggr\}.
\label{eq:thgtlm}
\end{eqnarray}

Excluding $\nabla_k$ terms, we obtain
\begin{equation}
\begin{aligned}
\tilde{\chi}^{THG}_{SSH}(\omega,\omega,\omega)
=\chi_{0}^{(3)}\frac{45}{1024}\int_{1}^{1/\delta}\frac{dx}
{\sqrt{(1-\delta^{2}x^2)(x^2-1)}}
\left\{\frac{1}{x^8(x^2-z^2)}-\frac{9}{x^8(x^2-(3z)^2)}\right\}
\label{eq:dkthgssh}
\end{aligned}\end{equation}

As $\delta \rightarrow 0$ in the Eq.\eqref{eq:dkefiorssh}, we have
\begin{equation}
\label{eq:dkthgtlm}
\begin{aligned}
\tilde{\chi}^{THG}_{TLM}(\omega,\omega,\omega)
=\chi_{0}^{(3)}\frac{1}{82944z^8}\left\{5(729f(z)-f(3z))
-8(216z^4+300z^2+455)\right\}
\end{aligned}
\end{equation}

\section{Discussions}
\label{sec:discu}
\subsection{resonant and non-resonant features}
\begin{figure}
\centerline{
\epsfxsize=15cm \epsfbox{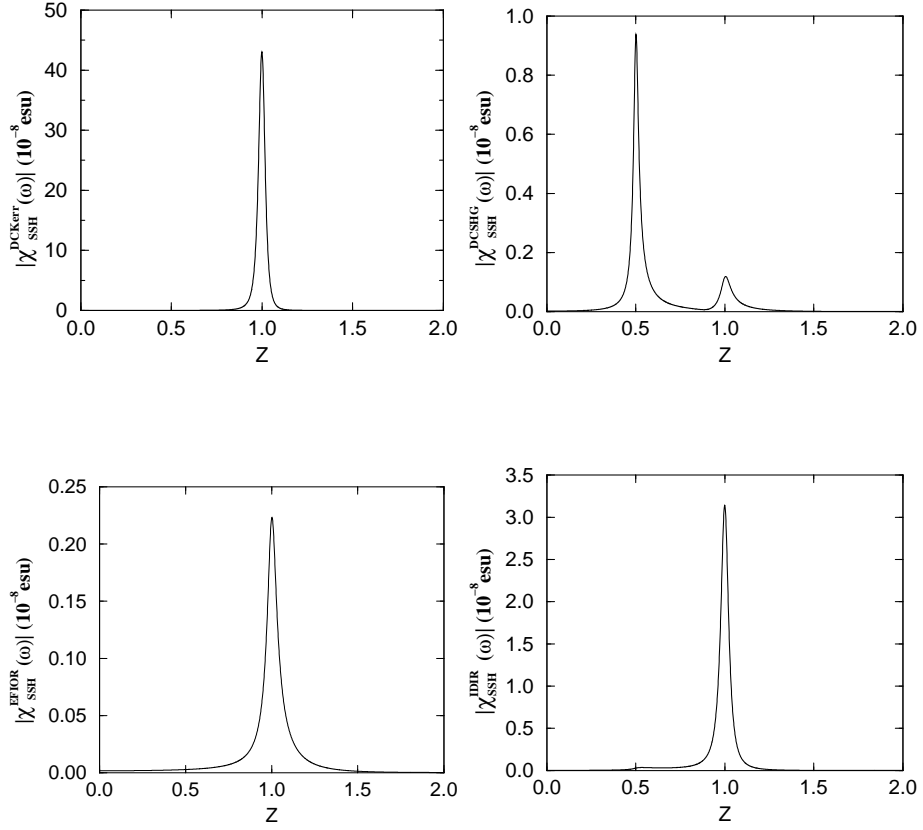}
}
\caption{The hyperpolarizabilities under the SSH model for frequency region 
$0<z<2.0$, where $z\equiv\hbar\omega/(2\Delta)$: DCKerr (top left), 
DCSHG (top right), IDIR (bottom left) and EFIOR (bottom right).}
\label{gr:sshhyper}
\end{figure}

Fig.\ref{gr:sshhyper} shows the hyperpolarizabilities of DCKerr, DCSHG, IDIR 
and EFIOR for SSH models. Choosing the same parameters in Section 
\ref{sec:hyper:fwm}, we see the magnitudes of the resonant peaks are in the 
following order: 
\begin{equation}
\begin{aligned}
\chi^{(3)}_{res}(-\omega;0,0,\omega)&>&
\chi^{(3)}_{res}(-\omega;\omega,-\omega,\omega)>
\chi^{(3)}_{res}(-2\omega;0,\omega,\omega)\\
&>&\chi^{(3)}_{res}(0;\omega;-\omega,0)>
\chi^{(3)}_{res}(-3\omega;\omega,\omega,\omega).
\end{aligned}
\label{eq:res}
\end{equation}

\begin{figure}
\centerline{
\epsfxsize=9cm \epsfbox{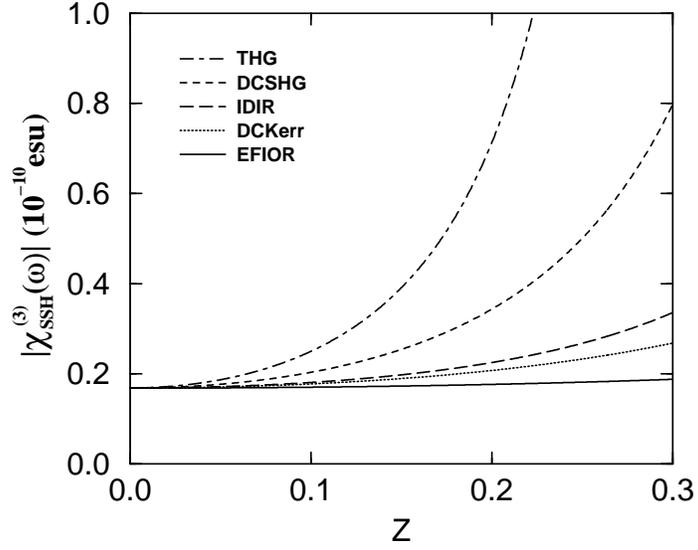}
}
\caption{The hyperpolarizabilities under the SSH model for non-resonant region:
EFIOR (real line), DCKerr (dotted line), IDIR (long dashed line), 
DCSHG (dashed line) and THG (dot-dashed line).}
\label{gr:fig_comp}
\end{figure}

Fig.\ref{gr:fig_comp} shows the comparison of hyperpolarizabilities of 
DCKerr, DCSHG, IDIR, EFIOR, THG for SSH models for low frequencies. 
Choosing the same parameters in Section 
\ref{sec:hyper}, we see the non-resonant features are in the following 
order:
\begin{equation}
\begin{aligned}
\chi^{(3)}_{non-res}(-3\omega;\omega,\omega,\omega)&>&
\chi^{(3)}_{non-res}(-2\omega;0,\omega,\omega)>
\chi^{(3)}_{non-res}(-\omega;\omega,-\omega,\omega)\\
&>&\chi^{(3)}_{non-res}(-\omega;0,0,\omega)> 
\chi^{(3)}_{non-res}(0;\omega,-\omega,0).
\end{aligned}
\label{eq:nonres}
\end{equation}

This non-resonant frequency dependence of nonlinear optical properties
is consistent with previous calculations for polyenes\cite{karna}.

\subsection{$\nabla_k$ term, Krammer-Kronig(KK) relation, the
overall permutation and Kleinman symmetries}
\label{sec:discu:dk}
There are some arguments about whether or not one should include the 
$\nabla_k$ terms in the calculations of nonlinear optical
properties. The $\nabla_k$ terms are usually considered to be related
to the intra-band current\cite{agrawal}. 
Otto suggested not to include this term 
because of the non-periodic property of $\nabla_k$\cite{otto1,otto2,otto3}, 
and thus the calculations would be purely based on the inter-band transition.
In this work, we compute the analytical results with or without $\nabla_k$ 
terms to show the differences of the results under both schemes. 

Considering the important physical contributions from the 
intra-band currents \cite{agrawal,kirtman}, we 
are in favor of including the $\nabla_k$ terms. Moreover, 
the restriction of our 
calculations in a unit cell implicitly imposes the periodic condition even for 
${\bf \nabla_k}$ operator. For the linear susceptibility $\chi^{(1)}$, 
our calculations\cite{mxu3} 
show that the $\nabla_k$ term make no actual contributions.
Due to the centro- or inversion symmetry for both SSH and TLM models,
the second-order susceptibility $\chi^{(2)}$ vanishes,
the first nonzero susceptibility is the third-order susceptibility 
$\chi^{(3)}$. From the formula, the $\nabla_k$ term causes the non-commuting
problem between operators in the third-order susceptibility $\chi^{(3)}$ 
calculations. Therefore, it breaks the overall permutation 
symmetry (between ($\om{1}$, $\om{2}$, $\om{3}$) and $\Omega=-\sum_i\om{i}$)
that is preserved in molecular systems\cite{butcher} where only bound states exist. 
Kleinman symmetry,  which is defined as the interchangeability of all
n indices in the rank n tensor $\chi^{(n)}$\cite{kleinman,butcher} and
derived from the overall permutation symmetry, 
is also broken even for low 
optical frequencies.
Our calculations clearly show the nonequivalent off-resonant behavior between 
$\chi^{DCKerr}(-\omega;0,0,\omega)$ and $\chi^{EFIOR}(0;\omega,-\omega,0)$.
To provide a theoretical result that can be measured by experiments, we
also perform the calculation of $\chi^{(3)}(\om{};\om{},\om{},-3\om{})$, 
which is the overall permutation of the THG 
$\chi^{(3)}(-3\om{};\om{},\om{},\om{})$. We obtain:
\begin{equation}\label{eq:sshoverall}
\begin{aligned}
\chi_{SSH}^{(3)}(\om{}; \om{},\om{},-3\om{})
&=\chi_{0}^{(3)}\frac{15}{1024}\int_{1}^{1/\delta}\frac{dx}
{\sqrt{(1-\delta^{2}x^2)(x^2-1)}} \\
&\left\{\frac{3}{x^8(x^2-z^2)}-\frac{27}{x^8(x^2-9z^2)}\right.\\
&+8(1-\delta^{2}x^2)(x^2-1)
\left\{-\frac{74}{9}\frac{1}{x^8(x^2-z^2)}
-\frac{11}{3}\frac{1}{x^6(x^2-z^2)^2}\right.\\
&\left.\left.+\frac{512}{9}\frac{1}{x^8(x^2-4z^2)}
-54\frac{1}{x^8(x^2-9z^2)}
+63\frac{1}{x^6(x^2-9z^2)^2}\right\}\right\}
\end{aligned}\end{equation}
Letting $\delta \rightarrow 0$ in \eqref{eq:sshoverall}, we obtain the
result under the TLM model:
\begin{equation}\label{eq:tlmoverall}
\begin{aligned}
\chi_{TLM}^{(3)}(\om{};\om{},\om{},-3\om{})
&=\chi_{0}^{(3)}\frac{5}{1024z^8}\left\{\frac{5}{3}(40z^2-61)f(z)+
\frac{16}{3}(4z^2-1)f(2z)\right.\\
&\left.-\frac{1}{243}(1944z^2-241)f(3z)
+\frac{32}{243}(27z^4-30z^2+805)\right\} \\
&=\chi_{0}^{(3)}(\frac{5}{28}+\frac{80}{33}z^2+\frac{28500}{1001}z^4 +O(z^6))
\ \ \ (z\rightarrow 0).
\end{aligned}\end{equation}
On the other hand, analytical expression for THG under the TLM model
is given by the formula (see, for example, \cite{mxu1,mxu2}):
\begin{equation}
\begin{aligned}
\chi_{TLM}^{(3)}(-3\omega;\omega,\omega,\omega)
&=\chi_0^{(3)} \frac{45}{128} \, \Biggl\{ -\frac{14}{3 z^8}-\frac{4}{15 z^4}
+\frac{(37-24 z^2)}{8 z^8} f(z)+\frac{(1-8 z^2)}{24 z^8} f(3z) \Biggr\}\\
&= \chi_{0}^{(3)} (\frac{5}{28}+\frac{80}{11}z^2+\frac{98580}{1001}z^4 +O(z^6))
\ \ \ (z\rightarrow 0).
\label{eq:tlmthg}
\end{aligned}
\end{equation}

\begin{figure}
\centerline{
\epsfxsize=9cm \epsfbox{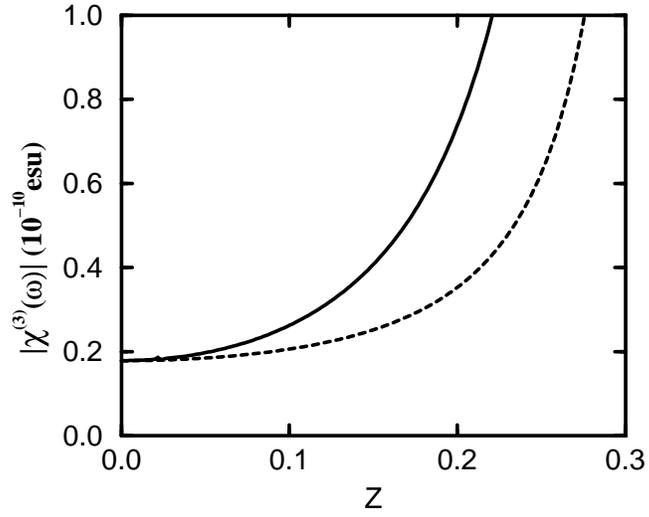}
}
\caption{The hyperpolarizabilities under the TLM model for non-resonant region:
$\chi^{(3)}_{TLM}(-3\om{};\om{},\om{},\om{})$ (real line) versus
$\chi^{(3)}_{TLM}(\om{};\om{},\om{},-3\om{})$ (long dashed line), where
$Z\equiv\hbar\om{}/2\Delta$.}
\label{gr:fig_kleinman}
\end{figure}

Both $\chi_{TLM}^{(3)}(\om{};\om{},\om{},-3\om{})$ and 
$\chi_{TLM}^{(3)}(-3\om{};\om{},\om{},\om{})$ in the non-resonant
region are plotted in Fig.\ref{gr:fig_kleinman}.
The numerical calculation shows 40\% difference between them 
when $z=1/6$ (corresponding to 0.3$eV$ or wavelength 4.14$\mu m$) in this 
model. Thus it clearly shows the break of overall permutation symmetry, while
the 
Kramers-Kronig relation is satisfied in Eq.\eqref{eq:tlmoverall} and
Eq.\eqref{eq:tlmthg}.
Subtracting two asymptotic expressions in Eq.\eqref{eq:tlmoverall} and
Eq.\eqref{eq:tlmthg}, we have the following relationship of the difference
in the off-resonant region:
\begin{equation}
\delta\chi^{(3)}(\omega) \propto \displaystyle \frac{e^4n_0t_0^3a^3
\hbar^2\omega^2}{\Delta^8}.
\label{eq:levine}
\end{equation}

By excluding the $\nabla_k$ terms, our calculations show both the overall 
permutation and Kleinman symmetries remain valid. For example, 
$\chi^{DCKerr}$=$\chi^{EFIOR}$ and $\chi^{(3)}(-3\om{};\om{},\om{},\om{})$=
$\chi^{(3)}(\om{};\om{},\om{},-3\om{})$ are preserved for all frequencies.
Obviously, it is the $\nabla_k$ term that breaks the overall
permutation in 
periodic 
systems. Though the experimental stimulated Raman scattering does not exhibit
the overall permutation symmetry for the resonant region\cite{dailey,butcher}, 
it seems that no experiment has been reported
to test the validity of the overall permutation symmetry for low frequency and
non-resonant region. Therefore, we suggest a new $\chi^{(3)}$ experiment
on infinite trans-polyacetylene chains to test the break of overall
permutation symmetry.

Recent experimental studies have already pointed out the deviation from the 
Kleinman symmetry for low optical frequencies in numerous
investigations of various optical systems\cite{dailey}. The assertion of
the general failure in Kleinman symmetry has been made in \cite{dailey}. 
Due to the similarity between the overall permutation and Kleinman
symmetry, Eq.\eqref{eq:levine} can also be used to explain the break
of Kleinman
symmetry in experiments qualitatively. Eq.\eqref{eq:levine} shows
that: (i). The break increases with decreasing band gap
and is proportional to $\omega^2$. These are consistent with the
previously reported experiment\cite{crane76};
(ii). The break increases with $t_0$ (the hopping of
$\pi$ electrons between the nearest-neighbor atoms). This explains the
experimental results that the deviation of Kleinman symmetry is favorable of
(20$\sim$50\%) the delocalized states such as aromatic
molecules\cite{wortmann92} and some
polymers or crystals\cite{lopez98,tsutsumi98},
while unfavorable of($\le$8\%) the
localized states such as molecular systems such as $O_2$, $N_2$,
etc.\cite{shelton85,shelton88}. On the other hand, the vanishing $\chi^{(2)}$
under the SSH or TLM model shows that some symmetries such as centro-symmetry,
etc, can suppress the deviation from Kleinman symmetry even for periodic
systems. This may explain why Kleinman symmetry is still preserved
in some $\chi^{(2)}$ experiments of crystals\cite{parsens71}.

\begin{figure}
\centerline{
\epsfxsize=15cm \epsfbox{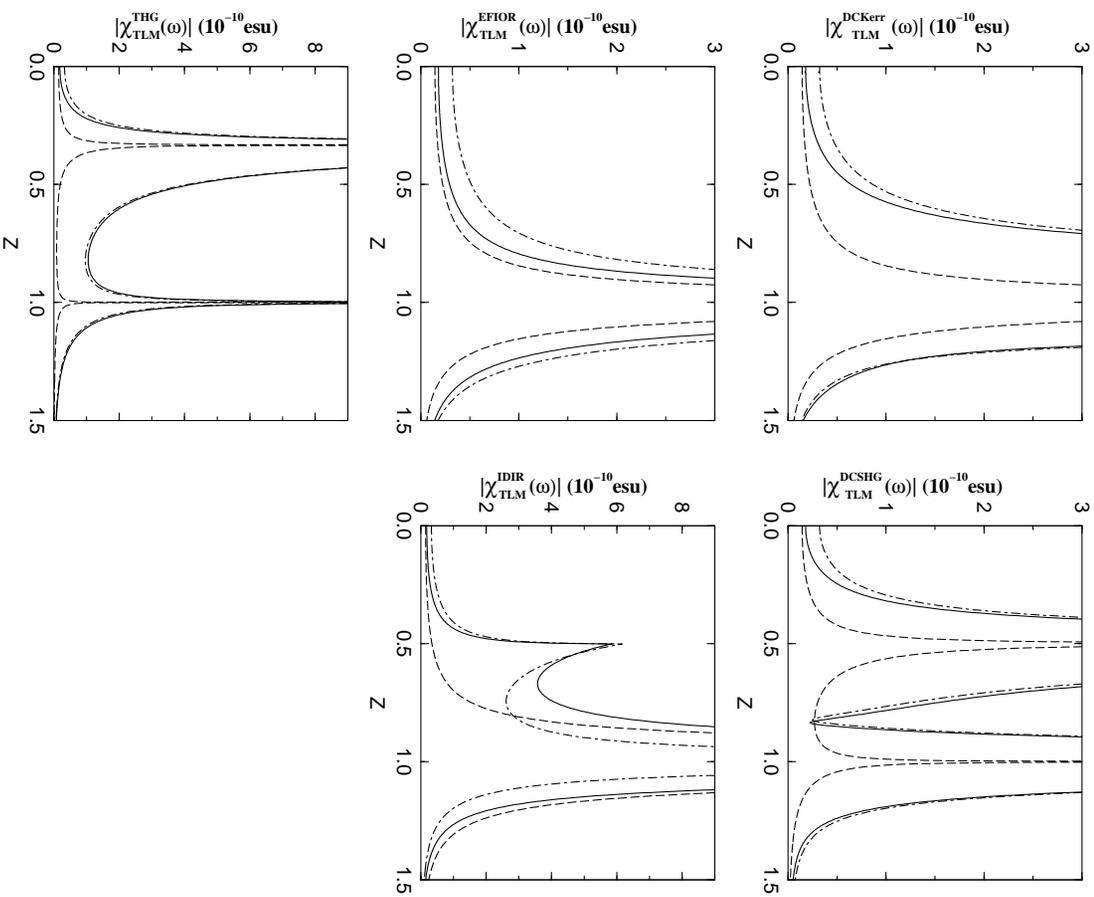}
}
\caption{The total(real line), intraband contribution(dot dashed line) and
interband contribution (long dashed line) of hyperpolarizabilities under 
TLM models - $\chi^{DCKerr}_{TLM}$ (top left), $\chi^{DCSHG}_{TLM}$ (top right),
$\chi^{EFIOR}_{TLM}$ (middle left), $\chi^{IDIR}_{TLM}$ (middle right) and 
$\chi^{THG}_{TLM}$ (bottom left).}
\label{gr:dkcompare}
\end{figure}

The magnitudes of the hyperpolarizabilities with $\nabla_k$ terms are
quite close to those 
without $\nabla_k$ terms in our results. In this sense, both results
can give the correct position of the resonant peaks qualitatively. However, 
we should notice that there is actually a sign difference
between the results of including and 
excluding $\nabla_k$ terms.
Therefore, the intraband current contribution to the hyperpolarizabilities 
is actually about as twice big as the interband contribution and can
not be neglected.
The total contribution (without $\nabla_k$ terms), interband 
contribution (with $\nabla_k$) and intraband contribution (the difference of
previous two) under TLM model are plotted in Fig.\ref{gr:dkcompare}.

\subsection{Zero frequency behaviors for hyperpolarizabilities}
Hyperpolarizabilities under zero frequency are also called static 
hyperpolarizabilities. To obtain the static hyperpolarizabilities, there 
are several different approaches. One way is to obtain the static 
polarizability by directly applying the static electronic field. For example,
by applying Resta's definition of dipole moment for periodic 
systems\cite{resta}, Soos and his coauthors obtained the polarizability of
one-dimensional Peierls-Hubbard model based on the static electronic
field\cite{soos3}. The other way is to obtain
the optical hyperpolarizabilities first, then let all frequencies approach 0. 
Obviously, both ways should yield the exactly same results of static
hyperpolarizabilities. Here we applies the latter method to do our 
calculations. 

We may use the results in Section \ref{sec:hyper} under TLM model to 
study the zero-frequency behaviors, with or without $\nabla_k$ terms. The 
zero-frequency results under SSH models shall be similar. 

{\bf 1. DCKerr}

Let $z\rightarrow 0$ in Eq.\eqref{eq:dckerrtlm}, we obtain
\begin{equation}
\chi^{DCKerr}_{TLM}(0,0,\omega\rightarrow 0)=\chi_{0}^{(3)}(\frac{5}{28}
+\frac{10}{11}z^2+ \frac{2160}{1001}z^4+\frac{1088}{273}z^6+O(z^8))
\end{equation}

Dropping $\nabla_k$ term and let $z\rightarrow 0$ in 
Eq.\eqref{eq:dkdckerrtlm}, we obtain
\begin{equation}
\tilde{\chi}^{DCKerr}_{TLM}(0,0,\omega\rightarrow 0)=\chi_{0}^{(3)}
(-\frac{1}{7}-\frac{50}{231}z^2-\frac{40}{143}z^4
-\frac{48}{143}z^6+O(z^8))
\end{equation}

{\bf 2. DCSHG}

Let $z\rightarrow 0$ in Eq.\eqref{eq:dcshgtlm}, we obtain
\begin{equation}
\chi^{DCSHG}_{TLM}(0,\omega\rightarrow 0,\omega\rightarrow 0)=\chi_{0}^{(3)}
(\frac{5}{28}+\frac{254}{77}z^2+\frac{68620}{3003}z^4+
\frac{375824}{3003}z^6+O(z^8))
\end{equation}

Dropping $\nabla_k$ term and let $z\rightarrow 0$ in 
Eq.\eqref{eq:dkdcshgtlm}, we obtain
\begin{equation}
\tilde{\chi}^{DCSHG}_{TLM}(0,\omega\rightarrow 0,\omega\rightarrow 0)=
\chi_{0}^{(3)} (-\frac{1}{7}-\frac{50}{77}z^2-\frac{360}{143}z^4
-\frac{1360}{143}z^6+O(z^8)).
\end{equation}

{\bf 3. IDIR}

Let $z\rightarrow 0$ in Eq.\eqref{eq:idirtlm}, we obtain
\begin{equation}
\chi^{IDIR}_{TLM}(\omega\rightarrow 0,-\omega\rightarrow 0,\omega\rightarrow 0)
=\chi_{0}^{(3)} (\frac{5}{28}+\frac{40}{33}z^2+\frac{5300}{1001}z^4+
\frac{256}{13}z^6+O(z^8)).
\end{equation}

Dropping $\nabla_k$ term and let $z\rightarrow 0$ in 
Eq.\eqref{eq:dkidirtlm}, we obtain
\begin{equation}
\tilde{\chi}^{IDIR}_{TLM}(\omega\rightarrow 0,-\omega\rightarrow 0,
\omega\rightarrow 0)
=\chi_{0}^{(3)}(-\frac{1}{7}-\frac{100}{231}z^2-\frac{120}{143}z^4
-\frac{192}{143}z^6+O(z^8)).
\end{equation}

{\bf 4. EFIOR}

Let $z\rightarrow 0$ in Eq.\eqref{eq:efiortlm}, we obtain
\begin{equation}
\chi^{EFIOR}_{TLM}(0,-\omega\rightarrow 0,\omega\rightarrow 0)=\chi_{0}^{(3)}
(\frac{5}{28}+\frac{10}{33}z^2+\frac{60}{143}z^4+
\frac{48}{91}z^6+O(z^8)).
\end{equation}

Dropping $\nabla_k$ term and let $z\rightarrow 0$ in 
Eq.\eqref{eq:dkefiortlm}, we obtain
\begin{equation}
\tilde{\chi}^{EFIOR}_{TLM}(0,-\omega,\omega)=\chi_{0}^{(3)}
(-\frac{1}{7}-\frac{50}{231}z^2-\frac{40}{143}z^4
-\frac{48}{143}z^6+O(z^8)).
\end{equation}

{\bf 5. THG}

From Eq.(3.3) in our previous work\cite{mxu1}, let $z\rightarrow 0$, we have
\begin{equation}
\chi^{THG}_{TLM}(\omega\rightarrow 0,\omega\rightarrow 0,\omega\rightarrow 0)
=\chi_{0}^{(3)} (\frac{5}{28}+\frac{80}{11}z^2+\frac{98580}{1001}z^4
+\frac{96640}{91}z^6+O(z^8)).
\end{equation}

Dropping $\nabla_k$ term and let $z\rightarrow 0$ in 
Eq.\eqref{eq:dkthgtlm}, we obtain
\begin{equation}
\tilde{\chi}^{THG}_{TLM}(\omega\rightarrow 0,\omega\rightarrow 0,
\omega\rightarrow 0)=\chi_{0}^{(3)}
(-\frac{1}{7}-\frac{100}{77}z^2-\frac{120}{11}z^4
-\frac{13120}{143}z^6+O(z^8)).
\end{equation}

From the zero frequency behaviors, we can see the difference of results
between having and not having $\nabla_k$ terms. Although the magnitude of
hyperpolarizabilities at (0,0,0) is quite close, $5\chi^{(3)}_0/28$ vs.
$4\chi^{(3)}_0/28$.

\subsection{Comparison with other theoretical results}
For the nonlinear properties under single electron periodic models, 
Genkin-Mednis developed an approach\cite{genkin} that was later applied to 
polymer systems by Agrawal, {\it et al}\cite{agrawal}. In those works,
general formulae of nonlinear optical response were developed. BY applying the 
Genkin-Mednis approach to the 
SSH model, Wu and Sun obtained the analytical format of third-harmonic 
generation (THG)\cite{cwu1,cwu2} that yields the same results as those
under the static current-current($J_0J_0$) correlation\cite{wwu}. As we 
discussed previously\cite{mxu1,mxu2}, the 
general results obtained here are quite similar but different from the results
obtained before. On the THG problem, our results are qualitatively close to 
the numerical results obtained by Yu, {\it et al}\cite{su1,su2} and 
Shuai, {\it et al}\cite{shuai}'s works. The reason for 
those difference is caused technically by the treatments of $\nabla_k$ 
operator, but physically by the gauge phase factor\cite{mxu3} that we are 
going to discuss further in the subsequent paper\cite{jiang1}.

Recently Kirtman, Gu and Bishop extended Genkin-Mednis method to the fully
coupled perturbed Hartree-Fock theory in discussing the hyperpolarizabilities 
of infinite chains\cite{kirtman}. In our results, it is still an uncoupled
treatment because the fully-filled valence and empty conduction band structures
in both SSH and TLM models lead no difference in the final 
results\cite{kirtman}.

Finally, we would like to point out that the results under $DD$ correlation are 
also different from those under $J_0J_0$ correlation\cite{wwu,yuri}. More
details will be discussed in our subsequent paper\cite{jiang1}.

\subsection{Some suggested experiments}
Our calculations predict the break of overall permutation symmetry
of hyperpolarizabilities for periodic systems in off-resonant regions. Despite 
of the wide acceptance of overall permutation symmetry in off-resonant 
regions\cite{butcher}, the direct measurement of the above assertion does 
not appear to be present in the literature. In our previous
work\cite{mxu4}, we suggested
experimentalists perform off-resonant $\chi^{(3)}$ measurements on some 
centro-symmetric 1D
periodic systems such as {\it trans-}polyacetylene, etc to directly test the
break of overall permutation symmetry and Eq.\eqref{eq:levine}. 
Though our calculations are based on 1D periodic models, the results 
may be applied to some delocalized two-dimensional materials such as benzene 
ring structures where the periodicity is along the circle of ring. Therefore,
experimentalists can also perform some off-resonant $\chi^{(3)}$ experiments on
some symmetric aromatic molecules to directly
test the overall permutation symmetry of hyperpolarizabilities. 

Besides the break of overall permutation symmetry, the relationships between
different nonlinear optical processes such as Eq.\eqref{eq:res} and
Eq.\eqref{eq:nonres} can also be tested experimentally.
\section{Conclusions}
\label{sec:concl}
Analytical expressions of the general four-wave-mixing (FWM) for infinite 
chains under the SSH and TLM models are first derived
under the $DD$ correlation by the field theory.
The hyperpolarizabilities of DCKerr, DCSHG, IDIR and EFIOR are obtained by
either including or excluding $\nabla_k$ terms. Contrary to the linear
response, intraband contributions to the hyperpolarizabilities dominate
the final results if they are included. $\nabla_k$ term also breaks the
overall permutation and Kleinman symmetry in $\chi^{(3)}$ even for the 
low frequency and non-resonant regions. The reported deviations
from Kleinman symmetry in experiments are explained qualitatively in this
work. A new feasible off-resonant $\chi^{(3)}$ experiment is
suggested to test the break of overall permutation symmetry in infinite 1D 
periodic polymer chains with centro-symmetry such as trans-polyacetylene, etc.
For the infinite single electron periodic systems, our calculations show the
following trends for the various third-order nonlinear optical processes
in the non-resonant region:
$\chi^{(3)}_{non-res}(-3\omega;\omega,\omega,\omega)>
\chi^{(3)}_{non-res}(-2\omega;0,\omega,\omega)>
\chi^{(3)}_{non-res}(-\omega;\omega,-\omega,\omega)
>\chi^{(3)}_{non-res}(-\omega;0,0,\omega) \ge
\chi^{(3)}_{non-res}(0;\omega,-\omega,0)$,
and in the resonant region:
$\chi^{(3)}_{res}(-\omega;0,0,\omega)>
\chi^{(3)}_{res}(-\omega;\omega,-\omega,\omega)>
\chi^{(3)}_{res}(-2\omega;0,\omega,\omega)
>\chi^{(3)}_{res}(0;\omega;-\omega,0)>
\chi^{(3)}_{res}(-3\omega;\omega,\omega,\omega)$.
We look forward to the experimental testings on the above
theoretical results.

This analytical calculations of hyperpolarizabilities are tedious, but helpful 
to illustrate both single electron models and theoretical methodologies in
nonlinear calculations. Based on single electronic models, more 
sophisticated models like Hubbard model\cite{soos1,mazumdar,zhang}, 
electron-hole pair model\cite{soos1,soos2}, etc, may be studied by
applying similar
techniques developed here.

\appendix
\section{Some integrals} \label{ap:si}
Define
\begin{equation}
\label{eq:fz}
f(z):=\int_{1}^{\infty}\frac{dx}{(x^2-z^2)\sqrt{x^2-1}}
\equiv \left \{
\begin{array}{lr}
\displaystyle  {\arcsin (z)\over z \sqrt{1-z^2}}  &(z^2<1),\\
\\
\displaystyle  -{\cosh^{-1} (z)\over z\sqrt{z^2-1}}+\displaystyle
{i\pi \over 2 z\sqrt{z^2-1}} &\ \ (z^2>1).
\end{array}
\right.
\end{equation}

\begin{equation}\label{eq:lnz}
L(n,z):=\int_{1}^{\infty}\frac{dx}{x^{2n}(x^2-z^2)^{5-n}\sqrt{x^2-1}},
\end{equation}

\begin{equation}\label{eq:mnz}
M(n,z):=\int_{1}^{\infty}\frac{\sqrt{x^2-1}dx}{x^{2n}(x^2-z^2)^{5-n}}.
\end{equation}

Using Maple, we obtain

\begin{equation}
L(4,z)=\frac{1}{105z^8}\left(105f(z)-(48z^6+56z^4+70z^2+105)\right),
\end{equation}

\begin{equation}
L(3,z)=\frac{1}{30z^8(z^2-1)}\left((-120z^2+105)f(z)+(16z^6+24z^4+50z^2-105)\right),
\end{equation}

\begin{equation}\begin{aligned}
L(2,z)&=\frac{1}{24z^8(z^2-1)^2}\left((144z^4-240z^2+105)f(z)\right.\\
&-\left.(16z^6+40z^4-170z^2+105)\right),
\end{aligned}\end{equation}

\begin{equation}
M(4,z)=\frac{1}{105z^8}\left((105z^2-105)f(z)-(8z^6+14z^4+35z^2-105)\right)
\end{equation}

\begin{equation}
M(3,z)=\frac{1}{30z^8}\left((-90z^2+105)f(z)+(4z^4+20z^2-105)\right)
\end{equation}

\begin{equation}
M(2,z)=\frac{1}{24z^8(z^2-1)}\left((72z^4-180z^2+105)f(z)+(-8z^4+110z^2-105)\right)
\end{equation}

\begin{equation}\begin{aligned}
M(1,z)&=\frac{1}{48z^8(z^2-1)^2}\left((-48z^6+216z^4-270z^2+105)f(z)\right.\\
&+\left.(-92z^4+200z^2-105)\right)
\end{aligned}\end{equation}

\begin{equation}\begin{aligned}
M(0,z)&=\frac{1}{384z^8(z^2-1)^3}\left((-192z^6+432z^4-360z^2+105)f(z)\right.\\
&+\left.(48z^6-248z^4+290z^2-105)\right)
\end{aligned}\end{equation}

\section{The derivation of FWM under SSH model} \label{ap:fwm}
We're trying to calculate the following integral:
\begin{eqnarray}
{\mathcal{I}} (\om{1},\om{2},\om{3})
=-\ib Tr(\dk G(0)\dk G(\om{1})\dk G(\om{1}+\om{2})\dk G(\om{1}+\om{2}+\om{3})),
\end{eqnarray}
where
\[\dk := \beta(k) \sg{2}+i\frac{\partial}{\partial k}\]
and
\[G(s) := \frac{\omega-s+\om{k} \sg{3}}
{(\omega-s)^{2}-\om{k}^{2}+i \epsilon}.\]

Let's define \[p_{\pm}:=\frac{1}{2}(1\pm \sg{3}),\]
\[\ak := \beta(k)\sg{2},\]
\[\bk := i\frac{\partial}{\partial k},\]
\[g_{+}(s):=\frac{1}{\omega-s-\om{k}+i\epsilon},\]
\[g_{-}(s):=\frac{1}{\omega-s+\om{k}-i\epsilon}.\]
Then it's easy to prove that
\[G(s) = g_{+}(s)\pp+g_{-}(s)\prm,\]
\[\ak p_{\pm} = p_{\mp} \ak,\]
\[\bk p_{\pm} = p_{\pm} \bk.\]

\begin{equation}
\begin{aligned}
{\mathcal{I}}(\om{1},\om{2},\om{3})
&= I_1+I_2+I_3+I_4+I_5+I_6+I_7+I_8\\
&:=
 -\ib Tr(\ak G(0)\ak G(\om{1})\ak G(\om{1}+\om{2})\ak G(\om{1}+\om{2}+\om{3}))\\
&-\ib Tr(\ak G(0)\ak G(\om{1})\bk G(\om{1}+\om{2})\bk G(\om{1}+\om{2}+\om{3}))\\
&-\ib Tr(\ak G(0)\bk G(\om{1})\ak G(\om{1}+\om{2})\bk G(\om{1}+\om{2}+\om{3}))\\
&-\ib Tr(\ak G(0)\bk G(\om{1})\bk G(\om{1}+\om{2})\ak G(\om{1}+\om{2}+\om{3}))\\
&-\ib Tr(\bk G(0)\ak G(\om{1})\ak G(\om{1}+\om{2})\bk G(\om{1}+\om{2}+\om{3}))\\
&-\ib Tr(\bk G(0)\ak G(\om{1})\bk G(\om{1}+\om{2})\ak G(\om{1}+\om{2}+\om{3}))\\
&-\ib Tr(\bk G(0)\bk G(\om{1})\ak G(\om{1}+\om{2})\ak G(\om{1}+\om{2}+\om{3}))\\
&-\ib Tr(\bk G(0)\bk G(\om{1})\bk G(\om{1}+\om{2})\bk G(\om{1}+\om{2}+\om{3}))
\end{aligned}
\end{equation}

\begin{enumerate}

\item
\begin{equation}
\begin{aligned}
I_1&=-\ib Tr(\ak G(0)\ak G(\om{1})\ak G(\om{1}+\om{2})\ak G(\om{1}+\om{2}+\om{3}
))\\
&=-\beta^{4}(k)\ib \gp{0} \gm{\om{1}} \gp{\om{1}+\om{2}}
\gm{\om{1}+\om{2}+\om{3}} \\
&-\beta^{4}(k)\ib \gm{0} \gp{\om{1}} \gm{\om{1}+\om{2}}
\gp{\om{1}+\om{2}+\om{3}}\\
&= I_1(\om{k})-I_1(-\om{k}),
\end{aligned}
\end{equation}

where
\begin{equation}
I_1(\om{k})=
-\frac{\beta^{4}(k)(4\om{k}-\om{1}-\om{3})}
{(2\om{k}-\om{1})(2\om{k}+\om{2})(2\om{k}-\om{3})
(2\om{k}-\om{1}-\om{2}-\om{3})}.
\end{equation}

\item
\begin{equation}
\begin{aligned}
I_2&=-\ib Tr(\ak G(0)\ak G(\om{1})\bk G(\om{1}+\om{2})\bk G(\om{1}+\om{2}+\om{3}
))\\
&=\ib \btk\gp{0}\btk\gm{\om{1}}\pk\gm{\om{1}+\om{2}}\pk\gm{\om{1}+\om{2}+\om{3}}
\\
&+\ib
\btk\gm{0}\btk\gp{\om{1}}\pk\gp{\om{1}+\om{2}}\pk\gp{\om{1}+\om{2}+\om{3}}\\
&= I_2(\om{k})-I_2(-\om{k}),
\end{aligned}
\end{equation}

where
\begin{equation}
I_2(\om{k})=
-\frac{\beta^{2}(k)}{2\om{k}-\om{1}}
(\pk\frac{1}{\omega-\om{1}-\om{2}+\om{k}}
\pk\frac{1}{\omega-\om{1}-\om{2}-\om{3}+\om{k}})|_{\omega=\om{k}}.
\end{equation}

\item
\begin{equation}
\begin{aligned}
I_3&=-\ib Tr(\ak G(0)\bk G(\om{1})\ak G(\om{1}+\om{2})\bk G(\om{1}+\om{2}+\om{3}
))\\
&=\ib \btk\gp{0}\pk\gp{\om{1}}\btk\gm{\om{1}+\om{2}}\pk\gm{\om{1}+\om{2}+\om{3}}
\\
&+\ib \btk\gm{0}\pk\gm{\om{1}}\btk\gp{\om{1}+\om{2}}\pk\gp{\om{1}+\om{2}+\om{3}}
\\
&= I_3(\om{k})-I_3(-\om{k}),
\end{aligned}
\end{equation}
where
\begin{equation}
\begin{aligned}
I_3(\om{k})&=
\beta(k)\pk\frac{\beta(k)}{(\omega-\om{2}+\om{k})
(\omega-\om{1}-\om{2}+\om{k})}\pk\frac{1}{\omega-\om{1}-\om{2}-\om{3}+\om{k}}
|_{\omega=\om{k}}\\
&+\beta(k)\pk\frac{\beta(k)}{\omega-\om{2}+\om{k}}
\pk\frac{1}{(\omega-\om{2}-\om{3}+\om{k})
(\omega-\om{1}-\om{2}-\om{3}+\om{k})}|_{\omega=\om{k}}\\
&-\frac{\partial \omega_{k}}{\partial
k}\frac{\beta^{2}(k)}{2(2\om{k}-\om{2})}
\frac{1}{2\om{k}-\om{2}}\pk\frac{1}{(2\om{k}-\om{2}-\om{3})(2\om{k}-\om{1}
-\om{2}-\om{3})}\\
&-\frac{\partial \omega_{k}}{\partial
k}\frac{\beta^{2}(k)}{2(2\om{k}-\om{2})}
\frac{1}{(2\om{k}-\om{2})(2\om{k}-\om{1}-\om{2})}\pk\frac{1}
{2\om{k}-\om{1}-\om{2}-\om{3}}\\
&-\frac{\partial \omega_{k}}{\partial
k}\frac{\beta^{2}(k)}{2(2\om{k}-\om{2})}
\frac{1}{2\om{k}-\om{2}-\om{3}}\pk\frac{1}{(2\om{k}-\om{2}-\om{3})
(2\om{k}-\om{1}-\om{2}-\om{3})}\\
&-\frac{\partial \omega_{k}}{\partial
k}\frac{\beta^{2}(k)}{2(2\om{k}-\om{2})}
\frac{1}{(2\om{k}-\om{2}-\om{3})(2\om{k}-\om{1}-\om{2}-\om{3})}
\pk\frac{1}{2\om{k}-\om{2}-\om{3}}.
\end{aligned}
\end{equation}

\item
\begin{equation}
\begin{aligned}
I_4&=-\ib Tr(\ak G(0)\bk G(\om{1})\bk G(\om{1}+\om{2})\ak G(\om{1}+\om{2}+\om{3}
))\\
&=\ib \btk\gp{0}\pk\gp{\om{1}}\pk\gp{\om{1}+\om{2}}\btk\gm{\om{1}+\om{2}+\om{3}}
\\
&+\ib \btk\gm{0}\pk\gm{\om{1}}\pk\gm{\om{1}+\om{2}}\btk\gp{\om{1}+\om{2}+\om{3}}
\\
&=I_{4}(\om{k})-I_{4}(-\om{k}),
\end{aligned}
\end{equation}
where
\begin{equation}
\begin{aligned}
I_{4}(\om{k})&=
-\beta(k)\frac{\partial^{2}}{\partial k^{2}}\frac{\beta(k)}
{(2\om{k}-\om{1}-\om{2}-\om{3})(2\om{k}-\om{2}-\om{3})(2\om{k}-\om{3})}\\
&+\frac{\beta(k)}{2}\pk\frac{\beta(k)}{2\om{k}-\om{3}}\pk
\frac{1}{(2\om{k}-\om{1}-\om{2}-\om{3})(2\om{k}-\om{2}-\om{3})}\\
&+\frac{\beta(k)}{2}\pk\frac{\beta(k)}{(2\om{k}-\om{3})(2\om{k}-\om{2}-\om{3})}
\pk \frac{1}{2\om{k}-\om{1}-\om{2}-\om{3}}\\
&-\frac{\beta^{2}(k)}{2\om{k}-\om{3}}(\pk\frac{1}{\omega-\om{2}-\om{3}+\om{k}}
\pk\frac{1}{\omega-\om{1}-\om{2}-\om{3}+\om{k}})|_{\omega=\om{k}}.
\end{aligned}
\end{equation}

\item
\begin{equation}
\begin{aligned}
I_5&=-\ib Tr(\bk G(0)\ak G(\om{1})\ak G(\om{1}+\om{2})\bk G(\om{1}+\om{2}+\om{3}
))\\
&=\ib \pk\gm{0}\btk\gp{\om{1}}\btk\gm{\om{1}+\om{2}}\pk\gm{\om{1}+\om{2}+\om{3}}
\\
&+\ib \pk\gp{0}\btk\gm{\om{1}}\btk\gp{\om{1}+\om{2}}\pk\gp{\om{1}+\om{2}+\om{3}}
\\
&=I_{5}(\om{k})-I_{5}(-\om{k}),
\end{aligned}
\end{equation}
where
\begin{equation}
\begin{aligned}
I_{5}(\om{k})=
-\pk\frac{\beta^{2}(k)}{2(2\om{k}+\om{1})(2\om{k}-\om{2})}
\pk\frac{1}{2\om{k}-\om{2}-\om{3}}.
\end{aligned}
\end{equation}

\item
\begin{equation}
\begin{aligned}
I_6&=-\ib Tr(\bk G(0)\ak G(\om{1})\bk G(\om{1}+\om{2})\ak G(\om{1}+\om{2}+\om{3}
))\\
&=\ib \pk\gm{0}\btk\gp{\om{1}}\pk\gp{\om{1}+\om{2}}\btk\gm{\om{1}+\om{2}+\om{3}}
\\
&+\ib \pk\gp{0}\btk\gm{\om{1}}\pk\gm{\om{1}+\om{2}}\btk\gp{\om{1}+\om{2}+\om{3}}
\\
&=I_{6}(\om{k})-I_{6}(-\om{k}),
\end{aligned}
\end{equation}
where
\begin{equation}
\begin{aligned}
I_{6}(\om{k})
&=\pk\frac{\beta(k)}{(2\om{k}+\om{1})(2\om{k}+\om{1}+\om{2})}
\pk\frac{\beta(k)}{2\om{k}-\om{3}}\\
&+\pk\frac{\beta(k)}{2\om{k}+\om{1}}
\pk\frac{\beta(k)}{(2\om{k}-\om{2}-\om{3})(2\om{k}-\om{3})}\\
&-\pk\frac{\beta^{2}(k)}{2(2\om{k}+\om{1})(2\om{k}-\om{3})}
\pk\frac{1}{2\om{k}-\om{2}-\om{3}}\\
&+\pk\frac{\beta^{2}(k)}{2(2\om{k}+\om{1})(2\om{k}-\om{3})}
\pk\frac{1}{2\om{k}+\om{1}+\om{2}}.
\end{aligned}
\end{equation}

\item
\begin{equation}
\begin{aligned}
I_7&=-\ib Tr(\bk G(0)\bk G(\om{1})\ak G(\om{1}+\om{2})\ak G(\om{1}+\om{2}+\om{3}
))\\
&=\ib \pk\gm{0}\pk\gm{\om{1}}\btk\gp{\om{1}+\om{2}}\btk\gm{\om{1}+\om{2}+\om{3}}
\\
&+\ib \pk\gp{0}\pk\gp{\om{1}}\btk\gm{\om{1}+\om{2}}\btk\gp{\om{1}+\om{2}+\om{3}}
\\
&=I_{7}(\om{k})-I_{7}(-\om{k}),
\end{aligned}
\end{equation}
where
\begin{equation}
\begin{aligned}
I_{7}(\om{k})
&=-\pk\frac{\beta^{2}(k)}{2(2\om{k}+\om{2})(2\om{k}-\om{3})}
\pk\frac{1}{2\om{k}+\om{1}+\om{2}}\\
&-\pk\frac{1}{2\om{k}+\om{1}+\om{2}}\pk
\frac{\beta^{2}(k)}{(2\om{k}+\om{2})(2\om{k}-\om{3})}.
\end{aligned}
\end{equation}

\item
\begin{equation}
\begin{aligned}
I_8&=-\ib Tr(\bk G(0)\bk G(\om{1})\bk G(\om{1}+\om{2})\bk G(\om{1}+\om{2}+\om{3}))\\
&=-\ib \bk\gp{0}\bk\gp{\om{1}}\bk\gp{\om{1}+\om{2}}\bk\gp{\om{1}+\om{2}+\om{3}}\\
&-\ib \bk\gm{0}\bk\gm{\om{1}}\bk\gm{\om{1}+\om{2}}\bk\gm{\om{1}+\om{2}+\om{3}}\\
&=0
\end{aligned}
\end{equation}
\end{enumerate}

The expression for ${\mathcal{I}}(\om{1},\om{2},\om{3})$ is extremely
messy. However, the quantity which we're really interested in is the
sum of its six permutations
\begin{equation}\begin{aligned}
{\mathcal{S}}&=\frac{1}{6}({\mathcal{I}}(\om{1},\om{2},\om{3})
+{\mathcal{I}}(\om{1},\om{3},\om{2})
+{\mathcal{I}}(\om{2},\om{1},\om{3})\\
&+{\mathcal{I}}(\om{2},\om{3},\om{1})
+{\mathcal{I}}(\om{3},\om{1},\om{2})
+{\mathcal{I}}(\om{3},\om{2},\om{1}))
\end{aligned}\end{equation}

Due to its highly symmetric character, we expect a great simplification
would occur. This is indeed the case as shown below. The key technique used
is that we can permute $\om{1}$, $\om{2}$ and $\om{3}$ freely when
computing the summation.
We present some details of the simplification procedure in the
following.
\begin{enumerate}

\item
\begin{equation}\begin{aligned}
I_5(\om{k})+I_6(\om{k})+I_7(\om{k})
&=&-\pk\frac{\beta^{2}(k)}{2(2\om{k}+\om{1})(2\om{k}-\om{2})}
\pk\frac{1}{2\om{k}-\om{2}-\om{3}}\\
&&+\pk\frac{\beta(k)}{(2\om{k}+\om{1})(2\om{k}+\om{1}+\om{2})}
\pk\frac{\beta(k)}{2\om{k}-\om{3}}\\
&&+\pk\frac{\beta(k)}{2\om{k}+\om{1}}
\pk\frac{\beta(k)}{(2\om{k}-\om{2}-\om{3})(2\om{k}-\om{3})}\\
&&-\pk\frac{\beta^{2}(k)}{2(2\om{k}+\om{1})(2\om{k}-\om{3})}
\pk\frac{1}{2\om{k}-\om{2}-\om{3}}\\
&&+\pk\frac{\beta^{2}(k)}{2(2\om{k}+\om{1})(2\om{k}-\om{3})}
\pk\frac{1}{2\om{k}+\om{1}+\om{2}}\\
&&-\pk\frac{\beta^{2}(k)}{2(2\om{k}+\om{2})(2\om{k}-\om{3})}
\pk\frac{1}{2\om{k}+\om{1}+\om{2}}\\
&&-\pk\frac{1}{2\om{k}+\om{1}+\om{2}}\pk
\frac{\beta^{2}(k)}{(2\om{k}+\om{2})(2\om{k}-\om{3})}\\
&=&-\pk\frac{\beta^{2}(k)}{(2\om{k}+\om{1})(2\om{k}-\om{2})}
\pk\frac{1}{2\om{k}-\om{2}-\om{3}}\\
&&+\pk\frac{\beta(k)}{(2\om{k}+\om{1})(2\om{k}+\om{1}+\om{2})}
\pk\frac{\beta(k)}{2\om{k}-\om{3}}\\
&&+\pk\frac{\beta(k)}{2\om{k}+\om{1}}
\pk\frac{\beta(k)}{(2\om{k}-\om{2}-\om{3})(2\om{k}-\om{3})}\\
&&-\pk\frac{1}{2\om{k}+\om{1}+\om{2}}\pk
\frac{\beta^{2}(k)}{(2\om{k}+\om{2})(2\om{k}-\om{3})}\\
&=&+\pk\frac{\beta(k)}{(2\om{k}+\om{1})(2\om{k}-\om{2}-\om{3})}
\pk\frac{\beta(k)}{2\om{k}-\om{2}}\\
&&-\pk\frac{\beta(k)}{(2\om{k}-\om{3})(2\om{k}+\om{1}+\om{2})}
\pk\frac{\beta(k)}{2\om{k}+\om{1}}\\
&=&+\pk\frac{\beta(k)}{(2\om{k}+\om{3})(2\om{k}-\om{1}-\om{2})}
\pk\frac{\beta(k)}{2\om{k}-\om{1}}\\
&&-\pk\frac{\beta(k)}{(2\om{k}-\om{3})(2\om{k}+\om{1}+\om{2})}
\pk\frac{\beta(k)}{2\om{k}+\om{1}}.
\end{aligned}\end{equation}

Since the above expression is an even function of $\om{k}$, we obtain
\begin{equation}
I_5+I_6+I_7 = 0.
\end{equation}

\item
\begin{equation}\begin{aligned}
I_4(\om{k})&=
-\frac{\beta(k)}{2}\pk\frac{1}{(2\om{k}-\om{1}-\om{2})
(2\om{k}-\om{1}-\om{2}-\om{3})}\pk\frac{\beta(k)}{2\om{k}-\om{1}}\\
&-\frac{\beta(k)}{2}\pk\frac{1}{2\om{k}-\om{1}-\om{2}-\om{3}}
\pk\frac{\beta(k)}{(2\om{k}-\om{1})(2\om{k}-\om{1}-\om{2})}\\
&-\frac{\beta^{2}(k)}{2\om{k}-\om{1}}(\pk\frac{1}{\omega-\om{1}-\om{2}+\om{k}}
\pk\frac{1}{\omega-\om{1}-\om{2}-\om{3}+\om{k}})|_{\omega=\om{k}}.
\end{aligned}\end{equation}

\item
Let
\begin{equation}
T(\om{k})=I_{2}(\om{k})+I_{3}(\om{k})+I_{4}(\om{k}).
\end{equation}
Then
\begin{equation}\begin{aligned}
T(\om{k})
&=
-\frac{\beta(k)}{2}\pk\frac{1}{(2\om{k}-\om{1}-\om{2})
(2\om{k}-\om{1}-\om{2}-\om{3})}\pk\frac{\beta(k)}{2\om{k}-\om{1}}\\
&-\frac{\beta(k)}{2}\pk\frac{1}{2\om{k}-\om{1}-\om{2}-\om{3}}
\pk\frac{\beta(k)}{(2\om{k}-\om{1})(2\om{k}-\om{1}-\om{2})}\\
&-\frac{2\beta^{2}(k)}{2\om{k}-\om{1}}
(\pk\frac{1}{\omega-\om{1}-\om{2}+\om{k}}
\pk\frac{1}{\omega-\om{1}-\om{2}-\om{3}+\om{k}})|_{\omega=\om{k}}\\
&+\beta(k)\pk\frac{\beta(k)}{(\omega-\om{1}+\om{k})
(\omega-\om{1}-\om{2}+\om{k})}\pk\frac{1}{\omega-\om{1}-\om{2}-\om{3}+\om{k}}
|_{\omega=\om{k}}\\
&+\beta(k)\pk\frac{\beta(k)}{\omega-\om{1}+\om{k}}
\pk\frac{1}{(\omega-\om{1}-\om{2}+\om{k})
(\omega-\om{1}-\om{2}-\om{3}+\om{k})}|_{\omega=\om{k}}\\
&-\frac{\partial \omega_{k}}{\partial k}
\frac{\beta^{2}(k)}{2(2\om{k}-\om{1})^{2}}
\pk\frac{1}{(2\om{k}-\om{1}-\om{2})(2\om{k}-\om{1}-\om{2}-\om{3})}\\
&-\frac{\partial \omega_{k}}{\partial
k}\frac{\beta^{2}(k)}{2(2\om{k}-\om{1})^{2}}
\frac{1}{(2\om{k}-\om{1}-\om{2})}\pk\frac{1}
{2\om{k}-\om{1}-\om{2}-\om{3}}\\
&-\frac{\partial \omega_{k}}{\partial
k}\frac{\beta^{2}(k)}{2(2\om{k}-\om{1})}
\frac{1}{(2\om{k}-\om{1}-\om{2})^{2}}\pk\frac{1}{
(2\om{k}-\om{1}-\om{2}-\om{3})}\\
&-\frac{\partial \omega_{k}}{\partial
k}\frac{\beta^{2}(k)}{(2\om{k}-\om{1})}
\frac{1}{(2\om{k}-\om{1}-\om{2})(2\om{k}-\om{1}-\om{2}-\om{3})}
\pk\frac{1}{2\om{k}-\om{1}-\om{2}}\\
&:=T_1+T_2+T_3+T_4+T_5+T_6+T_7+T_8+T_9,
\end{aligned}\end{equation}
where
\begin{eqnarray*}
T_3+T_4+T_5&=&
+\beta(k)\left(\pk\frac{\beta(k)}{\omega-\om{1}+\om{k}}\right)
\left(\frac{1}{\omega-\om{1}-\om{2}+\om{k}}\pk
\frac{1}{\omega-\om{1}-\om{2}-\om{3}+\om{k}}\right)
|_{\omega=\om{k}}\\
&&+\beta(k)\left(\pk\frac{\beta(k)}{\omega-\om{1}+\om{k}}\right)
\left(\pk\frac{1}{(\omega-\om{1}-\om{2}+\om{k})
(\omega-\om{1}-\om{2}-\om{3}+\om{k})}\right)
|_{\omega=\om{k}}\\
&&+\frac{\beta^{2}(k)}{2\om{k}-\om{1}}\pk\frac{1}
{\omega-\om{1}-\om{2}-\om{3}+\om{k}}
\pk\frac{1}{\omega-\om{1}-\om{2}+\om{k}}|_{\omega=\om{k}}\\
&=&+\frac{\partial \beta(k)}{\partial k}
\frac{\beta(k)}{2(2\om{k}-\om{1})(2\om{k}-\om{1}-\om{2})}
\pk
\frac{1}{2\om{k}-\om{1}-\om{2}-\om{3}}\\
&&+\frac{\partial \beta(k)}{\partial k}
\frac{\beta(k)}{2(2\om{k}-\om{1})}
\pk
\frac{1}{(2\om{k}-\om{1}-\om{2})(2\om{k}-\om{1}-\om{2}-\om{3})}\\
&&-\frac{\partial \om{k}}{\partial k}
\frac{\beta^{2}(k)}{2(2\om{k}-\om{1})^{2}}
\frac{1}{(2\om{k}-\om{1}-\om{2})}
\pk
\frac{1}{2\om{k}-\om{1}-\om{2}-\om{3}}\\
&&-\frac{\partial \om{k}}{\partial k}
\frac{\beta^{2}(k)}{2(2\om{k}-\om{1})^{2}}
\pk
\frac{1}{(2\om{k}-\om{1}-\om{2})(2\om{k}-\om{1}-\om{2}-\om{3})}\\
&&+\frac{\beta^{2}(k)}{4(2\om{k}-\om{1})}\left(\pk
\frac{1}{2\om{k}-\om{1}-\om{2}}\right)
\left(\pk\frac{1}{2\om{k}-\om{1}-\om{2}-\om{3}}\right)\\
&&+\frac{\beta^{2}(k)}{2(2\om{k}-\om{1})}
\frac{1}{(2\om{k}-\om{1}-\om{2}-\om{3})}
\frac{\partial^{2}}{\partial k^{2}}
\frac{1}{2\om{k}-\om{1}-\om{2}}\\
&&+\frac{\partial \om{k}}{\partial k}
\frac{\beta^{2}(k)}{(2\om{k}-\om{1})}
\frac{1}{(2\om{k}-\om{1}-\om{2})(2\om{k}-\om{1}-\om{2}-\om{3})}
\pk\frac{1}{2\om{k}-\om{1}-\om{2}}.
\end{eqnarray*}

Then
\begin{eqnarray*}
T'&=&T_3+T_4+T_5+T_6+T_7+T_8+T_9\\
&=&+\frac{\partial \beta(k)}{\partial k}
\frac{\beta(k)}{2(2\om{k}-\om{1})(2\om{k}-\om{1}-\om{2})}
\pk
\frac{1}{2\om{k}-\om{1}-\om{2}-\om{3}}\\
&&+\frac{\partial \beta(k)}{\partial k}
\frac{\beta(k)}{2(2\om{k}-\om{1})}
\pk
\frac{1}{(2\om{k}-\om{1}-\om{2})(2\om{k}-\om{1}-\om{2}-\om{3})}\\
&&+\frac{\beta^{2}(k)}{2}
\left(\pk\frac{1}{2\om{k}-\om{1}}\right)
\left(
\frac{1}{2\om{k}-\om{1}-\om{2}}
\pk
\frac{1}{2\om{k}-\om{1}-\om{2}-\om{3}}\right)\\
&&+\frac{\beta^{2}(k)}{2}
\left(\pk\frac{1}{2\om{k}-\om{1}}\right)
\left(
\pk
\frac{1}{(2\om{k}-\om{1}-\om{2})(2\om{k}-\om{1}-\om{2}-\om{3})}\right)\\
&&+\frac{\beta^{2}(k)}{2(2\om{k}-\om{1})}\left(\pk
\frac{1}{2\om{k}-\om{1}-\om{2}}\right)
\left(\pk\frac{1}{2\om{k}-\om{1}-\om{2}-\om{3}}\right)\\
&&+\frac{\beta^{2}(k)}{2(2\om{k}-\om{1})}
\frac{1}{(2\om{k}-\om{1}-\om{2}-\om{3})}
\frac{\partial^{2}}{\partial k^{2}}
\frac{1}{2\om{k}-\om{1}-\om{2}}\\
&=&+\frac{\beta(k)}{2}
\left(\pk\frac{\beta(k)}{2\om{k}-\om{1}}\right)
\left(
\frac{1}{2\om{k}-\om{1}-\om{2}}
\pk
\frac{1}{2\om{k}-\om{1}-\om{2}-\om{3}}\right)\\
&&+\frac{\beta(k)}{2}
\left(\pk\frac{\beta(k)}{2\om{k}-\om{1}}\right)
\left(
\pk
\frac{1}{(2\om{k}-\om{1}-\om{2})(2\om{k}-\om{1}-\om{2}-\om{3})}\right)\\
&&+\frac{\beta^{2}(k)}{2(2\om{k}-\om{1})}
\pk\frac{1}{2\om{k}-\om{1}-\om{2}-\om{3}}
\pk\frac{1}{2\om{k}-\om{1}-\om{2}}.
\end{eqnarray*}
Therefore
\begin{eqnarray*}
T(\om{k})&=& T_1+T_2+T'\\
&=&-\frac{\beta(k)}{2}
\frac{1}{(2\om{k}-\om{1}-\om{2})(2\om{k}-\om{1}-\om{2}-\om{3})}
\frac{\partial^2}{\partial k^2}
\frac{\beta(k)}{2\om{k}-\om{1}}\\
&&-\frac{\beta(k)}{2}
\frac{1}{2\om{k}-\om{1}-\om{2}-\om{3}}
\frac{\partial^2}{\partial k^2}
\frac{\beta(k)}{(2\om{k}-\om{1})(2\om{k}-\om{1}-\om{2})}\\
&&+\frac{\beta^{2}(k)}{2(2\om{k}-\om{1})(2\om{k}-\om{1}-\om{2}-\om{3})}
\frac{\partial^2}{\partial k^2}
\frac{1}{2\om{k}-\om{1}-\om{2}}\\
&=&-\frac{\beta(k)}{(2\om{k}-\om{1}-\om{2})(2\om{k}-\om{1}-\om{2}-\om{3})}
\frac{\partial^2}{\partial k^2}
\frac{\beta(k)}{2\om{k}-\om{1}}\\
&&-\frac{\beta(k)}{2\om{k}-\om{1}-\om{2}-\om{3}}
\left(\pk\frac{\beta(k)}{2\om{k}-\om{1}}\right)
\left(\pk\frac{1}{2\om{k}-\om{1}-\om{2}}\right)\\
&=&-\frac{\beta(k)}{2\om{k}-\om{1}-\om{2}-\om{3}}
\pk\frac{1}{2\om{k}-\om{1}-\om{2}}\pk\frac{\beta(k)}{2\om{k}-\om{1}}.
\end{eqnarray*}
\end{enumerate}

Combining the above results, we obtain
\begin{eqnarray*}
{\mathcal{I}}'(\om{1},\om{2},\om{3})&=&I_1(\om{k})+T(\om{k})
-I_1(-\om{k})-T(-\om{k})\\
&=&
-\frac{\beta^{4}(k)(4\om{k}-\om{1}-\om{3})}
{(2\om{k}-\om{1})(2\om{k}+\om{2})(2\om{k}-\om{3})
(2\om{k}-\om{1}-\om{2}-\om{3})}\\
&&-\frac{\beta^{4}(k)(4\om{k}+\om{1}+\om{3})}
{(2\om{k}+\om{1})(2\om{k}-\om{2})(2\om{k}+\om{3})
(2\om{k}+\om{1}+\om{2}+\om{3})}\\
&&-\frac{\beta(k)}{2\om{k}-\om{1}-\om{2}-\om{3}}
\pk\frac{1}{2\om{k}-\om{1}-\om{2}}\pk\frac{\beta(k)}{2\om{k}-\om{1}}\\
&&-\frac{\beta(k)}{2\om{k}+\om{1}+\om{2}+\om{3}}
\pk\frac{1}{2\om{k}+\om{1}+\om{2}}\pk\frac{\beta(k)}{2\om{k}+\om{1}},
\end{eqnarray*}
and
\begin{equation}
{\mathcal{S}} = \frac{1}{6} \sum_{{\mathcal{P}}(\om{1},\om{2},\om{3})}
{\mathcal{I}}'(\om{1},\om{2},\om{3})
\end{equation}

We now proceed to compute the general four-wave-mixing third order
susceptibility $\chi^{(3)}(\om{1},\om{2},\om{3})$ which is defined by:
\begin{equation}
\chi^{(3)}(\om{1},\om{2},\om{3}) =
\frac{2e^{4}n_{0}}{\hbar^3}\frac{1}{L}\sum_{k}{\mathcal{S}}.
\end{equation}
For infinite chains, the summation is replaced by the integral:
\begin{equation}\label{eq:chi3}
\begin{aligned}
\chi^{(3)}(\om{1},\om{2},\om{3}) &=
\frac{2e^{4}n_{0}}{\hbar^3}\int_{-\frac{\pi}{2a}}^{\frac{\pi}{2a}}{\mathcal{S}}
\frac{dk}{2\pi}
=\frac{2e^{4}n_{0}}{\pi\hbar^3}\int_{0}^{\frac{\pi}{2a}}{\mathcal{S}}dk\\
&=\frac{1}{6}\sum_{{\mathcal{P}}(\om{1},\om{2},\om{3})}
\frac{2e^{4}n_{0}}{\pi\hbar^3}\int_{0}^{\frac{\pi}{2a}}
{\mathcal{I}}'(\om{1},\om{2},\om{3})dk\\
&:=\sum_{{\mathcal{P}}(\om{1},\om{2},\om{3})}\chi(\om{1},\om{2},\om{3}).
\end{aligned}
\end{equation}

We first recall:
\begin{equation}
\om{k}=\frac{\epsilon(k)}{\hbar},
\end{equation}

\begin{equation}
\epsilon(k)=\sqrt{[2t_{0}\cos{(ka)}]^2+[\Delta\sin{(ka)}]^2},
\end{equation}

\begin{equation}
\beta(k)=-\frac{\Delta t_{0}a}{\epsilon^{2}(k)}.
\end{equation}
Furthermore, we define the following quantities:
\begin{equation}
\delta=\frac{\Delta}{2t_{0}}.
\end{equation}

\begin{equation}
\chi_{0}^{(3)}=\frac{8}{45}\frac{e^4n_{0}}{\pi\Delta^3}\frac{a^3}{\delta^3},
\end{equation}

\begin{equation}
z_{i}=\frac{\hbar\om{i}}{2\Delta}, \quad \text{for $i=1,2,3$},
\end{equation}

Let us introduce change of variable:
\begin{equation}
x=\frac{\hbar\om{k}}{\Delta}.
\end{equation}
Then for $0<k<\frac{\pi}{2a}$, we have:
\begin{equation}\label{dxdk}
\frac{dx}{dk}=-\frac{a}{\delta}\frac{\sqrt{(1-\delta^{2}x^2)(x^2-1)}}{x},
\end{equation}
and hence

\begin{equation}\begin{aligned}
\chi(\om{1},\om{2},\om{3})
&=-\chi_{0}^{(3)}\frac{15}{1024}\int_{1}^{1/\delta}\frac{xdx}
{\sqrt{(1-\delta^{2}x^2)(x^2-1)}} \\
&\{\frac{(2x-z_{1}-z_{3})}
{x^{8}(x-z_{1})(x+z_{2})(x-z_{3})(x-z_{1}-z_{2}-z_{3})}\\
&+\frac{(2x+z_{1}+z_{3})}
{x^{8}(x+z_{1})(x-z_{2})(x+z_{3})(x+z_{1}+z_{2}+z_{3})}\\
&+\left(\frac{2\delta}{a}\right)^2\frac{1}{x^{2}(x-z_{1}-z_{2}-z_{3})}
\pk\frac{1}{x-z_{1}-z_{2}}\pk\frac{1}{x^{2}(x-z_{1})}\\
&+\left(\frac{2\delta}{a}\right)^2\frac{1}{x^{2}(x+z_{1}+z_{2}+z_{3})}
\pk\frac{1}{x+z_{1}+z_{2}}\pk\frac{1}{x^{2}(x+z_{1})}\}\\
&=\chi_{0}^{(3)}\frac{15}{1024}\int_{1}^{1/\delta}\frac{xdx}
{\sqrt{(1-\delta^{2}x^2)(x^2-1)}} \\
&\{-\frac{(2x-z_{1}-z_{3})}
{x^{8}(x-z_{1})(x+z_{2})(x-z_{3})(x-z_{1}-z_{2}-z_{3})}\\
&-\frac{(2x+z_{1}+z_{3})}
{x^{8}(x+z_{1})(x-z_{2})(x+z_{3})(x+z_{1}+z_{2}+z_{3})}\\
&+\left(\frac{2\delta}{a}\right)^2
\frac{1}{x-z_{1}-z_{2}}\left(\pk\frac{1}{x^{2}(x-z_{1})}\right)
\left(\pk\frac{1}{x^{2}(x-z_{1}-z_{2}-z_{3})}\right)\\
&+\left(\frac{2\delta}{a}\right)^2
\frac{1}{x+z_{1}+z_{2}}\left(\pk\frac{1}{x^{2}(x+z_{1})}\right)
\left(\pk\frac{1}{x^{2}(x+z_{1}+z_{2}+z_{3})}\right)\}
\end{aligned}\end{equation}

By \eqref{dxdk}, we have
\begin{equation}\label{eq:chi}
\begin{aligned}
\chi(\om{1},\om{2},\om{3})
&=\chi_{0}^{(3)} \frac{15}{1024} \int_{1}^{1/\delta}\frac{xdx}
{\sqrt{(1-\delta^{2}x^2)(x^2-1)}} \\
&\left\{-\frac{(2x-z_{1}-z_{3})}
{x^{8}(x-z_{1})(x+z_{2})(x-z_{3})(x-z_{1}-z_{2}-z_{3})}\right.\\
&-\frac{(2x+z_{1}+z_{3})}
{x^{8}(x+z_{1})(x-z_{2})(x+z_{3})(x+z_{1}+z_{2}+z_{3})}\\
&+\frac{4(1-\delta^{2}x^2)(x^2-1)}{x^2(x-z_{1}-z_{2})}
\left(\px\frac{1}{x^{2}(x-z_{1})}\right)
\left(\px\frac{1}{x^{2}(x-z_{1}-z_{2}-z_{3})}\right)\\
&+\left.\frac{4(1-\delta^{2}x^2)(x^2-1)}{x^2(x+z_{1}+z_{2})}
\left(\px\frac{1}{x^{2}(x+z_{1})}\right)
\left(\px\frac{1}{x^{2}(x+z_{1}+z_{2}+z_{3})}\right)\right\}\\
&=\chi_{0}^{(3)} \frac{15}{1024} \int_{1}^{1/\delta}\frac{xdx}
{\sqrt{(1-\delta^{2}x^2)(x^2-1)}} \\
&\left\{-\frac{(2x-z_{1}-z_{3})}
{x^{8}(x-z_{1})(x+z_{2})(x-z_{3})(x-z_{1}-z_{2}-z_{3})}\right.\\
&-\frac{(2x+z_{1}+z_{3})}
{x^{8}(x+z_{1})(x-z_{2})(x+z_{3})(x+z_{1}+z_{2}+z_{3})}\\
&+\frac{4(1-\delta^{2}x^2)(x^2-1)}{x^8(x-z_{1}-z_{2})}
\frac{(3x-2z_{1})(3x-2(z_{1}+z_{2}+z_{3}))}
{(x-z_{1})^{2}(x-z_{1}-z_{2}-z_{3})^2}\\
&+\left.\frac{4(1-\delta^{2}x^2)(x^2-1)}{x^8(x+z_{1}+z_{2})}
\frac{(3x+2z_{1})(3x+2(z_{1}+z_{2}+z_{3}))}
{(x+z_{1})^{2}(x+z_{1}+z_{2}+z_{3})^2}\right\}\\
&:=I_{1}(\om{1},\om{2},\om{3})+I_{2}(\om{1},\om{2},\om{3})
+I_{3}(\om{1},\om{2},\om{3})+I_{4}(\om{1},\om{2},\om{3}),
\end{aligned}\end{equation}

With the aid of Mathematica, we find that
\begin{equation}\begin{aligned}
J&=\sum_{{\mathcal{P}}(\om{1},\om{2},\om{3})}(I_{1}(\om{1},\om{2},\om{3})+
I_{2}(\om{1},\om{2},\om{3}))\\
&=\frac{15 \chi_{0}^{(3)}}{128(z_{1}+z_{2})(z_{2}+z_{3})(z_{3}+z_{1})}
\int_{1}^{1/\delta}\frac{dx}{\sqrt{(1-\delta^{2}x^2)(x^2-1)}}\\
&\frac{1}{x^8}\left(\sum_{i=1}^{3}\frac{z_{i}^{3}}{x^{2}-z_{i}^{2}}-
\frac{(z_{1}+z_{2}+z_{3})^3}{x^{2}-(z_{1}+z_{2}+z_{3})^2}\right)
\end{aligned}\end{equation}

and

\begin{equation}\begin{aligned}
K&=\sum_{{\mathcal{P}}(\om{1},\om{2},\om{3})}(I_{3}(\om{1},\om{2},\om{3})
+I_{4}(\om{1},\om{2},\om{3}))\\
&=\frac{15\chi_{0}^{(3)}}{128}\int_{1}^{1/\delta}\frac{\sqrt{(1-\delta^{2}x^2)
(x^2-1)}dx}{x^8}\\
&\sum_{i=1}^{3}\frac{(z_{1}+z_{2})^{5}(z_{1}+z_{2}-2z_{3})}
{z_{1}^{2}z_{2}^{2}z_{3}^{2}(x^{2}-(z_{1}+z_{2})^{2})}\\
&+\frac{-(z_{1}+z_{2}+z_{3})^{3}(
(z_{1}+z_{2}+z_{3})^{4}(z_{1}z_{2}+z_{2}z_{3}+z_{3}z_{1})
-z_{1}z_{2}z_{3}(8(z_{1}+z_{2}+z_{3})^{3}+7z_{1}z_{2}z_{3}))}
{z_{1}^{2}z_{2}^{2}z_{3}^{2}(z_{1}+z_{2})(z_{2}+z_{3})(z_{3}+z_{1})
(x^{2}-(z_{1}+z_{2}+z_{3})^{2})}\\
&+\frac{(z_{1}+z_{2}+z_{3})^{3}((z_{1}+z_{2}+z_{3})^{3}+z_{1}z_{2}z_{3})
(x^{2}+(z_{1}+z_{2}+z_{3})^{2})}
{z_{1}z_{2}z_{3}(z_{1}+z_{2})(z_{2}+z_{3})(z_{3}+z_{1})
(x^{2}-(z_{1}+z_{2}+z_{3})^{2})^{2}}\\
&+\sum_{i=1}^{3}\frac{-z_{1}^{2}(z_{1}-2(z_{2}+z_{3}))(x^{2}+z_{1}^{2})}
{z_{2}z_{3}(z_{2}+z_{3})(x^{2}-z_{1}^{2})^{2}}\\
&+\sum_{i=1}^{3}\frac{-z_{1}^{2}(z_{1}^{2}(z_{2}+z_{3})
-8z_{2}z_{3}(z_{2}+z_{3})-2z_{1}(z_{2}+z_{3})^{2}+7z_{1}z_{2}z_{3})}
{z_{2}^{2}z_{3}^{2}(z_{2}+z_{3})(x^{2}-z_{1}^{2})}
\end{aligned}\end{equation}

Using integration by parts, we obtain
\begin{equation}\begin{aligned}
L&=\int_{1}^{1/\delta}\frac{\sqrt{(1-\delta^{2}x^2)(x^2-1)}dx}{x^8}
\frac{x^{2}+z^{2}}{(x^{2}-z^{2})^{2}}\\
&=\int_{1}^{1/\delta}\frac{\sqrt{(1-\delta^{2}x^2)(x^2-1)}dx}{x^8}
\frac{1}{x^2-z^2}\left(-8+\frac{x^2(1+\delta^2-2\delta^2 x^2)}
{(1-\delta^{2}x^2)(x^2-1)}\right)\\
&=\int_{1}^{1/\delta}\frac{\sqrt{(1-\delta^{2}x^2)(x^2-1)}dx}{x^8}
\frac{1}{x^2-z^2}\left(-7+\frac{1-\delta^2 x^4}
{(1-\delta^{2}x^2)(x^2-1)}\right)
\end{aligned}\end{equation}

Therefore,
\begin{equation}\begin{aligned}
K&=\frac{15\chi_{0}^{(3)}}{128}\int_{1}^{1/\delta}\frac{\sqrt{(1-\delta^{2}x^2)
(x^2-1)}dx}{x^8}\\
&\sum_{i=1}^{3}\frac{(z_{1}+z_{2})^{5}(z_{1}+z_{2}-2z_{3})}
{z_{1}^{2}z_{2}^{2}z_{3}^{2}(x^{2}-(z_{1}+z_{2})^{2})}\\
&+\frac{-(z_{1}+z_{2}+z_{3})^{6}}{z_{1}^{2}z_{2}^{2}z_{3}^{2}
(x^{2}-(z_{1}+z_{2}+z_{3})^{2})}\\
&+\frac{(z_{1}+z_{2}+z_{3})^{3}((z_{1}+z_{2}+z_{3})^{3}+z_{1}z_{2}z_{3})}
{z_{1}z_{2}z_{3}(z_{1}+z_{2})(z_{2}+z_{3})(z_{3}+z_{1})}
\frac{1-\delta^2 x^4}
{(x^{2}-(z_{1}+z_{2}+z_{3})^{2})(1-\delta^{2}x^2)(x^2-1)}\\
&+\sum_{i=1}^{3}\frac{-z_{1}^{2}(z_{1}-2(z_{2}+z_{3}))}
{z_{2}z_{3}(z_{2}+z_{3})(x^{2}-z_{1}^{2})}
\frac{1-\delta^2 x^4}{(1-\delta^{2}x^2)(x^2-1)}\\
&+\sum_{i=1}^{3}\frac{-z_{1}^{2}(z_{1}^{2}
-2z_{1}(z_{2}+z_{3})+6z_{2}z_{3})}
{z_{2}^{2}z_{3}^{2}(x^{2}-z_{1}^{2})}
\end{aligned}\end{equation}

Finally,
\begin{equation}\begin{aligned}
\chi^{(3)} &=J+K\\
&=\frac{15\chi_{0}^{(3)}}{128}\int_{1}^{1/\delta}\frac{dx}
{x^8 \sqrt{(1-\delta^{2}x^2)(x^2-1)}}\\
&\left\{\frac{Z^{3}}
{z_{1}z_{2}z_{3}\sigma}
\frac{Z^{3}-(Z^{3}+z_{1}z_{2}z_{3})
\delta^2 x^4}{x^{2}-Z^{2}}
+\sum_{i=1}^{3}\frac{z_{i}^{3}}{\sigma (x^{2}-z_{i}^{2})}
+\sum_{P(z_1,z_2,z_3)}\frac{z_{1}^{2}(-z_{1}+2(z_{2}+z_{3}))(1-\delta^{2}x^{4})}
{2z_{2}z_{3}(z_{2}+z_{3})(x^{2}-z_{1}^2)}\right\}\\
&+\frac{15\chi_{0}^{(3)}}{128}\int_{1}^{1/\delta}
\frac{\sqrt{(1-\delta^{2}x^2)(x^2-1)}dx}{x^8}\\
&\left\{\sum_{P(z_1,z_2,z_3)}\left(\frac{(z_{1}+z_{2})^{5}(z_{1}+z_{2}-2z_{3})}
{2z_{1}^{2}z_{2}^{2}z_{3}^{2}(x^{2}-(z_{1}+z_{2})^{2})}
+\frac{-z_{1}^{2}(z_{1}^{2}
-2z_{1}(z_{2}+z_{3})+6z_{2}z_{3})}
{2z_{2}^{2}z_{3}^{2}(x^{2}-z_{1}^{2})}\right)
-\frac{Z^{6}}{z_{1}^{2}z_{2}^{2}z_{3}^{2}
(x^{2}-Z^{2})}\right\}
\end{aligned}\end{equation}

where
\begin{equation}
\sigma := (z_{1}+z_{2})(z_{2}+z_{3})(z_{3}+z_{1}),
\end{equation}
and
\begin{equation}
Z := z_1+z_2+z_3.
\end{equation}

We now consider the case $\delta\rightarrow0$.

Using the results in Appendix \ref{ap:si}, we obtain
\begin{equation}\begin{split}
\chi^{(3)} &= \frac{15\chi_{0}^{(3)}}{128}
\left\{\frac{Z^{6}}
{z_{1}z_{2}z_{3}\sigma}L(4,Z)-\frac{Z^{6}}{z_{1}^{2}z_{2}^{2}z_{3}^{2}}M(4,Z)
+\sum_{i=1}^{3}\frac{z_{i}^{3}}{\sigma}L(4,z_{i})\right.\\
&+\left.\sum_{P(z_1,z_2,z_3)}\frac{z_{1}^{2}(-z_{1}+2(z_{2}+z_{3}))}
{2z_{2}z_{3}(z_{2}+z_{3})}L(4,z_{1})\right\} \\
&+\frac{15\chi_{0}^{(3)}}{128}
\left\{\sum_{P(z_1,z_2,z_3)}\frac{(z_{1}+z_{2})^{5}(z_{1}+z_{2}-2z_{3})}
{2z_{1}^{2}z_{2}^{2}z_{3}^{2}}M(4,z_{1}+z_{2})\right.\\
&+\left.\frac{-z_{1}^{2}(z_{1}^{2}
-2z_{1}(z_{2}+z_{3})+6z_{2}z_{3})}
{2z_{2}^{2}z_{3}^{2}}M(4,z_{1})\right\}
\end{split}\end{equation}

\bibliography{}

\begin{figure}
\centerline{
\epsfxsize=9cm \epsfbox{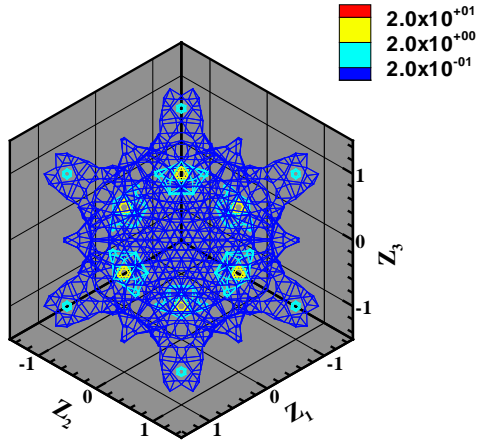}
}
\caption{(Color) The magnitude of four-wave-mixing(FMW) 
$\chi^{(3)}(-\Omega;\om{1},\om{2},\om{3})$ under SSH model
is in unit of $10^{-8}$esu. $Z_i$ is defined by Eq.\eqref{eq:zi}.}
\label{gr:fwm}
\end{figure}

\end{document}